\documentclass[journal]{IEEEtran}
\ifCLASSINFOpdf
\else
\fi

\usepackage{amsmath,amsfonts,bm}
\usepackage[amsmath,thmmarks,framed]{ntheorem}
\usepackage{framed}
\usepackage{color}
\definecolor{gray}{rgb}{0.6,0.6,0.6}

\newframedtheorem{defn}{Definition}
\newframedtheorem{thm}{Theorem}
\newframedtheorem{prop}{Proposition}

\usepackage{algorithm}
\usepackage{algorithmic}
\usepackage{multirow}

\usepackage{booktabs}
\usepackage{graphicx}
\usepackage{upgreek}
\usepackage{graphics}
\usepackage{subfig}
\usepackage{mathrsfs}
\usepackage{amssymb}
\newcommand*{\QEDA}{\hfill\ensuremath{\blacksquare}}

\usepackage{url}
\usepackage{hyperref}
\usepackage{cite}
\makeatletter

\newcommand{\Rmnum}[1]{\expandafter \@slowromancap\romannumeral #1@}
\makeatother
\begin{document}
%
\title{Reconstruction of Missing Big Sensor Data}
%
%
%

\author{Yongshuai Shao and Zhe Chen
\thanks{Yongshuai Shao and Zhe Chen are with the Department of Computer Engineering, School of Computer Science and Engineering, Northeastern University, Shenyang, Liaoning, P. R. China. E-mail: chenzhe@mail.neu.edu.cn, zchen42@students.tntech.edu.}}%
\maketitle

\begin{abstract}
  With ubiquitous sensors continuously monitoring and collecting large amounts of information, there is no doubt that this is an era of big data. One of the important sources for scientific big data is the datasets collected by Internet of things (IoT). It's considered that these datesets contain highly useful and valuable information. For an IoT application to analyze big sensor data, it is necessary that the data are clean and lossless. However, due to unreliable wireless link or hardware failure in the nodes, data loss in IoT is very common. To reconstruct the missing big sensor data, firstly, we propose an algorithm based on matrix rank-minimization method. Then, we consider IoT with multiple types of sensor in each node. Accounting for possible correlations among multiple-attribute sensor data, we propose tensor-based methods to estimate missing values. Moreover, effective solutions are proposed using the alternating direction method of multipliers. Finally, we evaluate the approaches using two real sensor datasets with two missing data-patterns, i.e., random missing pattern and consecutive missing pattern. The experiments with real-world sensor data show the effectiveness of the proposed methods.
\end{abstract}

\begin{IEEEkeywords}
 Big Sensor Data, Data Reconstruction, Rank Minimization, Tensor Completion, ADMM.
\end{IEEEkeywords}

%
\IEEEpeerreviewmaketitle

\section{Introduction}
\IEEEPARstart{W}{ith} ubiquitous sensors continuously monitoring and collecting large amounts of information, there is no doubt this is an era of big data. One of the important sources for scientific big data is the datasets collected by Internet of things (IoT)~\cite{1,2}. It's considered that these datesets contain highly useful and valuable information.
Learning from these large amounts of sensor data is expected to bring significant science advances and improvements in quality of life. For example, scientists understand the demand for plant evolution based on light condition in forest~\cite{3}, discover the eruption omen by monitoring the shake of the volcano~\cite{4}, and provide valuable information to individual by analyzing the data relevant to the health of patient~\cite{5}. All of these scientific research work heavily depends on the accuracy of original sensor data.

However, due to unreliable wireless link or hardware failure in the nodes, big data from sensors are often subject to corruption and losses. Furthermore, missing data become larger as sensor networks grow in scale \cite{6}. These missing values cause great difficulties for data analysis methods such as classification, prediction, and other machine learning methods, which often fail to deal with missing values, especially when the amount of missing data is very large. Therefore, in order to better analyze big sensor data for IoT applications, it is necessary that the received data are clean and lossless. As a result, it is urgent and important to design effective methods to reconstruct missing values in big sensor data.
\subsection{Existing Approaches and Their Limitations}
A great deal of existing work has devoted to predict missing sensor data. Most techniques are based on temporal methods, spatial methods, or spatial-temporal methods.
Temporal methods include last seen~\cite{7} and linear interpolation. These methods leverage temporal correlations among readings in the same node.
K-Nearest-Neighbor (KNN)~\cite{8} is a classical local interpolation method. KNN simply utilizes the values of the nearest K neighbors to estimate the missing one. Window association rule mining (WARM)~\cite{9} and freshness association rule mining (FARM)~\cite{10} study the estimation of missing data based on association rules among spatially-correlated neighbors. These techniques belong to spatial-correlation based missing-data estimation methods.
A general method for reconstruction of missing data is suggested in~\cite{11}, which exploits both temporal and spatial redundancy to characterize the phenomenon being monitored. This method considers model hierarchies and selects a proper model from state-space or input/output linear and nonlinear models for each sensor, which may result in great complexity when dealing with massive sensors.
In data estimation using statistical model (DESM)~\cite{12}, a missing reading is predicted using the linear combination of the previous reading of the sensor and the current reading of the neighboring sensor, weighted by the Pearson correlation between the two sensors.
The applying k-nearest neighbor estimation (AKE) method~\cite{13} adopts the linear regression model to describe the spatial correlation of sensor data among different sensor nodes and utilizes the data information of all neighboring nodes to estimate the missing data. These techniques belong to spatial-temporal correlation based missing-data estimation methods.

The above methods may suffer from over-relying on assumptions about data. For example, when the sensors have a long-time sampling interval, the usefulness of temporal methods may drop rapidly as the number of consecutively missing reading increases. Besides, using spatial correlation can often lead to worse estimation results as non-existent correlations are imposed between nearby sensors. Moreover, some sensor datasets favor spatial correlation over temporal correlation or vice versa. In general, on one hand, such assumptions may not hold for various datasets. On the other hand, the exactness of assumption based models directly affects the accuracy of prediction results. So it is necessary to find a new way to learn latent structures from sensor data without heavily relying on such a priori knowledge.

Recently, the intrinsic low-rank property of high-dimensional data has been considered.
In contrast to many existing approaches that make strong assumption about data, Li et al. applies matrix factorization (MF)-based method to recover missing data~\cite{14}, which learns inter-sensor and intra-sensor correlations by exploiting their latent similarity. Finally, they extend the methods to account for possible correlations among multiple types of sensors.
In paper~\cite{15,16,17,18}, matrix completion theory is used to recover the missing data in sink node for large-scale wireless sensor networks. However, their main work focuses on energy saving
by selecting a sample of entries from each node uniformly and randomly.
In paper~\cite{19,20,21,22}, compressive sensing technique is applied to the reconstruction of sensor data. A novel approach based on compressive sensing to reconstruct massive missing sensor data is proposed in~\cite{19}. By analyzing real sensor data, the features of spatial correlation, temporal stability, and low-rank structure are exhibited. A multiple-attributes-based recovery algorithm is proposed in paper~\cite{20}. This algorithm combines the benefits of compressive sensing and the correlation of attributes. In paper~\cite{21}, an algorithm combining the benefits of compressive sensing, spatial-temporal correlation, and multi-attribute correlation features is proposed. And a novel sensory data recovery algorithm is proposed in~\cite{22}, which exploits the spatial and temporal joint-sparse feature. However, to the best of our knowledge, little research has been conducted on multi-attribute sensor data reconstruction.

Other related work for estimating missing values is focused on tensor completion. Gandy et al.~\cite{23} applies the alternative direction method of multipliers algorithm (ADMM) to solve the tensor completion problem with Gaussian observation noise. And Liu et al.~\cite{25} proposes a high-accuracy low-rank tensor completion algorithm (HaLRTC) to solve the tensor completion problem without consideration of noises. However, both of them mainly estimate the missing data in visual datasets.
\subsection{Our Work and Contribution}
Based on the aforementioned existing related works, in this paper, we investigate the methods of reconstructing missing big sensor data. Our work is fourfold.

Firstly, based on the fact that most sensor data have low-rank structures, we utilize rank minimization technique to recover missing sensor data. In order to solve the rank minimization problem, we propose an ADMM-based rank minimization algorithm, namely ADRM. ADRM takes full advantage of the low-rank structure feature of real sensor data by computing the minimal low-rank approximations of the incomplete sensor data matrix.

Secondly, considering that nodes in IoT often have multiple sensor types and monitor multi-attribute data, we propose a tensor-based method to reconstruct the multi-attribute sensor data as well as an ADMM-based multi-attribute sensor data reconstruction algorithm, namely ADMAR, to reconstruct the big sensor data. ADMAR is based on the assumption that the constructed tensor sensor data is jointly low-rank in all modes.

Thirdly, considering that the constructed tensor through multi-attribute sensor data may not always be low-rank in all modes, we propose a relaxed version of ADMAR (R-ADMAR), which only requires that the tensor is low-rank in certain modes.

Finally, we evaluate the effectiveness of our proposed approaches using two real datasets. We study two patterns of missing data, i.e., the random missing pattern and the consecutive missing pattern. In the experiment of reconstructing single-attribute sensor data, we compare ADMR with the classical interior point method and KNN. It shows that ADMR performs the best with the aforementioned missing patterns. In the experiment of reconstructing tensor-based multi-attribute sensor data, we demonstrate that ADMAC outperforms the existing EM-based Tucker decomposition algorithm and shows a little advantage over HaLRTC. In addition, R-ADMAR performs better than ADMAR when the constructed tensor is low-rank only in certain modes.

Our contributions are summarized as follows.

Firstly, we use rank minimization technique to recover missing sensor data and propose an algorithm, named ADMR, based on ADMM method.

Secondly, to the best of our knowledge, this is the first work to apply tensor-based method to sensor data reconstruction problem.

Thirdly, we propose a tensor-based algorithm, named ADMAC, to reconstruct multi-attribute sensor data.

Finally, in order to overcome the shortcoming of ADMAC, we propose a relaxed version of the ADMAC algorithm, namely R-ADMAC.

The rest parts of the paper are organised as follows. In Section II, we formulate the big sensor data reconstruction problem. Section III proposes the algorithm for reconstructing single-attribute sensor data. Section IV proposes the algorithm for reconstructing multi-attribute sensor data. Section V proposes the R-ADMAC algorithm. The performance is evaluated in Section VI. And Section VII concludes this paper.

\section{Problem Formulation}

Suppose $n$ nodes are deployed in an area, each of which equips $k$ sensors to monitor different attributes at the same time. The monitoring period consists of $t$ time slots. The gathered sensor data in one node can be organized in the following format~\cite{20},
\begin{table}[htp]
\centering
\begin{tabular}{|c|c|c|c|c|}
\hline
  Sensor ID & Time Stamp & Attribute 1 & Attribute 2 & ...\\
\hline
\end{tabular}
\end{table}
where sensor ID stands for sensor identity number, time stamp represents sampling time, and the attributes can be temperature, humidity, and so on.

Firstly, we consider single-attribute sensor data reconstruction problem. Let $\mathbf{M}$ denote a matrix of sensor data with one attribute collected by $n$ node within $t$ time slots. Each $\mathbf{M}$ is an $n\times t$ matrix. Due to data loss in IoT, $\mathbf{M}$ is usually an incomplete matrix. The available information of $\mathbf{M}$ is a set of entries $m_{p,q}, (p,q) \in \Omega$, where $\Omega$ is the set of sampled entries of $\mathbf{M}$. This process is represented using a sampling operator $\mathscr{P}_{\Omega}(\cdot)$, which is defined by:
\begin{equation}
[\mathscr{P}_{\Omega}(\mathbf{X})]_{ij} = \left\{
     \begin{array}{lcl}
     {x_{i,j},  \quad \text{if}\, (i,j) \in \Omega}\\
     {0, \qquad \text{othewise.}}
     \end{array}
     \right.
\end{equation}
Therefore, the single-attribute sensor-data reconstruction problem can be defined as follows.

Given subsets of $\mathbf{M}$, which is denoted as $\mathscr{P}_{\Omega}(\mathbf{M})$, find an optimal solution denoted as $\hat{\mathbf{M}}$,
\begin{equation}
\begin{aligned}
  &\text{minimize} \quad \quad \quad \Vert \mathrm{\hat{\textbf{M}}}-\mathrm{\textbf{M}}\Vert_{F}\\
  &\text{subject to}  \quad \mathscr{P}_{\Omega}(\mathrm{\hat{\textbf{M}}})=\mathscr{P}_{\Omega}(\mathrm{\textbf{M}}),
\end{aligned}
\end{equation}
where $\Vert \cdot \Vert_{F}$ represents the Frobenius norm of matrix.

Then we consider the multi-attribute sensor-data reconstruction problem.
Here we get $k$ matrices of sensor data, $\mathbf{M}_{1},\mathbf{M}_{2},..,\mathbf{M}_{k}$, each of which denotes one attribute collected by $n$ node within $t$ time slots.
Due to data loss in IoT, the matrices we finally obtain are $\mathscr{P}_{\Omega_{i}}(\mathbf{M}_{i}), i=1,...,k$.
Our problem is to recover a series of raw data $\mathbf{M}_{1},...,\mathbf{M}_{k}$ from sampled incomplete matrices $\mathscr{P}_{\Omega_{1}}(\mathbf{M}_{1}),..., \mathscr{P}_{\Omega_{k}}(\mathbf{M}_{k})$ as precisely as possible.

In order to solve this problem we constitute a third-order tensor using the series of sampled incomplete matrices $\mathscr{P}_{\Omega_{1}}(\mathbf{M}_{1}),..., \mathscr{P}_{\Omega_{k}}(\mathbf{M}_{k})$. The three modes represent sensor time stamp $I_{t}$, sensor ID $I_{ID}$, and attributes $I_{a}$, respectively. Thus we finally obtain a tensor of multi-attribute sensor data, i.e., $\mathcal{T} \in R^{I_{t}\times I_{ID}\times I_{a}}$.
The multi-attribute sensor data reconstruction problem is defined as follows:

Given subsets of $\mathcal{T}$ denoted as $\mathscr{P}_{\Omega}(\mathcal{T})$, find an optimal solution $\hat{\mathcal{T}}$,
\begin{equation}
\begin{aligned}
  &\text{minimize} \quad \quad \quad \Vert \hat{\mathcal{T}}-\mathcal{T}\Vert_{F}\\
  &\text{subject to}  \quad \mathscr{P}_{\Omega}(\hat{\mathcal{T}})=\mathscr{P}_{\Omega}(\mathcal{T}),
\end{aligned}
\end{equation}
where $\Vert \cdot \Vert_{F}$ is the Frobenius norm of tensor.

\section{Reconstruction of Single-Attribute Sensor Data}
In this section, we propose an ADMM based rank minimization algorithm, namely ADRM, to address single-attribute sensor-data reconstruction problem. ADRM takes full advantage of the low-rank structure feature of real sensor data by computing the minimal low-rank approximations of the incomplete sensor data matrix.

\subsection{Matrix Rank-Minimization Based Approach}
Let $\mathbf{M}$ denote received single-attribute sensor data matrix, $\mathbf{M} \in R^{n\times t}$. Due to low rank structure feature, which has been revealed in many papers such as~\cite{15,19}, the missing values in $\mathbf{M}$ can be recovered using rank minimization.
\begin{equation}
\begin{aligned}
  &\mathop{\text{minimize}} \limits_{\mathbf{X}}  \qquad \quad rank(\mathbf{X}) \\
  &\text{subject to} \quad x_{i,j}=m_{i,j},(i,j)\in \Omega ,
\end{aligned}
\end{equation}
where the elements of $\mathbf{M}$ in the set $\Omega$ are given while the remaining elements are missing. $rank(\mathbf{X}$) denotes the rank of matrix $\mathbf{X}$. For the sake of simplicity, the constraint condition can be summarized using $\mathscr{P}_{\Omega}(\mathbf{X})=\mathscr{P}_{\Omega}(\mathbf{M})$.
However, the problem shown in Eq.~(4) is an NP-hard problem, hence it cannot be easily used in practice. A widely used alternative is the convex relaxation,
\begin{equation}
\begin{aligned}
  &\mathop{\text{minimize}} \limits_{\mathbf{X}}  \quad \quad \qquad \Vert \mathbf{X} \Vert_{*} \\
  &\text{subject to} \quad \; \mathscr{P}_{\Omega}(\mathbf{X})=\mathscr{P}_{\Omega}(\mathbf{M}),
\end{aligned}
\end{equation}
where $\Vert \mathbf{X}\Vert_{*}$ is the nuclear norm of matrix $\mathbf{X}$, that is, the sum of singular values of $\mathbf{X}$. The nuclear norm minimization problem is the general model of matrix completion~\cite{26}.

In practical terms, the noises in sensory data may lead to the over-fitting problem. Thus, we consider the following relaxed problem:
\begin{equation}
\begin{aligned}
  \mathop{\text{minimize}} \limits_{\mathbf{X}} \quad \Vert \mathbf{X} \Vert_{*} +(1/2\lambda)\Vert \mathscr{P}_{\Omega}(\mathbf{X})-\mathscr{P}_{\Omega}(\mathbf{M})\Vert_{F}^{2},
\end{aligned}
\end{equation}
where parameter $0<\lambda \le 1$, which controls the fit to the constraint $\mathscr{P}_{\Omega}(\mathbf{X})=\mathscr{P}_{\Omega}(\mathbf{M})$. Consider a continuation technique for decreasing the value of $\lambda$ towards convergence, the $\mathscr{P}_{\Omega}(\mathbf{X})$ is close to but not equal to $\mathscr{P}_{\Omega}(\mathbf{M})$.

For convenience, we define sampling matrix $\mathbf{B}$, where
\begin{equation}
b_{i,j} = \left\{
     \begin{array}{lcl}
     {1, \quad \text{if}\;(i,j) \in \Omega}\\
     {0, \quad \text{otherwise}.}
     \end{array}
     \right.
\end{equation}
Obviously, $\mathbf{B}$ is an $n\times t$ binary matrix and indicates whether data in $\mathbf{M}$ are missing or not.

We now define the single-attribute sensor-data reconstruction problem.
\begin{defn}
$\mathbf{B}$ is the sampling matrix and $\mathbf{M}$ is the incomplete sensor data matrix that is to be recovered. Then the missing values in $\mathbf{M}$ can be effectively estimated by solving the following convex optimization problem,
\begin{equation}
\begin{aligned}
  \mathop{\text{minimize}} \limits_{\mathbf{X}} \quad \Vert \mathbf{X} \Vert_{*} +(1/2\lambda)\Vert \mathbf{B}\cdot \mathbf{X}-\mathbf{B}\cdot \mathbf{M}\Vert_{F}^{2},
\end{aligned}
\end{equation}
where $(\cdot)$ denotes the element-wise production of matrix, $\lambda$ is a parameter.
\end{defn}

Problem (8) is a typical convex optimization problem. It can be transformed to a semidefinite programming problem and solved using interior point methods. In our experiment, in order to solve problem (8), we use CVX, a package for specifying and solving convex programs~\cite{27} as a contrast experiment.

Because CVX uses interior point methods to solve convex optimization, it has very high computation complexity. Especially when dealing with large-scale data, CVX takes much time or even cannot run. So, recently many first-order methods based algorithms have been proposed to solve convex optimization problems.

In this paper, we propose a matrix-rank minimization based algorithm, namely ADRM, based on ADMM method to solve problem (8).

\subsection{ADMM}
The ADMM
is a convex optimization algorithm dating back to the early 1980's. It has attracted attention again recently due to the fact that it is efficient to tackle large-scale problems and may be implemented in parallel and distributed computational environments.

The general ADMM model is expressed as follows~\cite{30}:
\begin{equation}
\begin{aligned}
  &\text{minimize} \quad f(\mathbf{x})+g(\mathbf{z}) \\
  &\text{subject to} \quad \mathbf{A}\mathbf{x} + \mathbf{B}\mathbf{z} = \mathbf{c},
\end{aligned}
\end{equation}
with variables $\mathbf{x} \in R^{n}$ and $\mathbf{z} \in R^{m}$, where $\mathbf{A} \in R^{p\times n}$, $\mathbf{B}\in R^{p\times m}$, and $\mathbf{c}\in R^{p}$. Assume that $f$ and $g$ are convex. By introducing a Lagrange multiplier $\mathbf{y}\in R^{p}$ to the equality constraint $\mathbf{A}\mathbf{x} + \mathbf{B}\mathbf{z} = \mathbf{c}$, we form the augmented Lagrangian function,
\begin{equation}
\begin{aligned}
 \mathcal{L}_{\rho}(\mathbf{x},\mathbf{z},\mathbf{y})&=f(\mathbf{x})+g(\mathbf{z})+\mathbf{y}^{T}(\mathbf{A}\mathbf{x} + \mathbf{B}\mathbf{z} - \mathbf{c})\\
 &+(\rho/2)\Vert \mathbf{A}\mathbf{x} + \mathbf{B}\mathbf{z} - \mathbf{c}\Vert_{2}^{2}.
\end{aligned}
\end{equation}
ADMM consists of the following iterations.
\begin{equation}
\begin{aligned}
 &\mathbf{x}^{k+1}=\mathop{\text{argmin}} \limits_{\mathbf{x}} \mathcal{L}_{\rho}(\mathbf{x},\mathbf{z}^{k},\mathbf{y}^{k})\\
 &\mathbf{z}^{k+1}=\mathop{\text{argmin}} \limits_{\mathbf{z}}\mathcal{L}_{\rho}(\mathbf{x}^{k+1},\mathbf{z},\mathbf{y}^{k})\\
 &\mathbf{y}^{k+1}=\mathbf{y}^{k}+\rho (\mathbf{A}\mathbf{x}^{k+1} + \mathbf{B}\mathbf{z}^{k+1} - \mathbf{c})
\end{aligned}
\end{equation}

The parameter $\rho$ can be any positive number. The convergence of the general ADMM algorithm is guaranteed~\cite{30} and the convergence rate is affected by $\rho$. In ADMM, $\mathbf{x}$ and $\mathbf{z}$ are updated in an alternating or sequential fashion, which accounts for the term \emph{alternating direction}.

\subsection{ADRM Algorithm for Reconstruction of Single-Attribute Sensor Data}
In order to apply the ADMM method to problem (8), we need to transform it into the ADMM form. We first rephrase (8) as below.

We introduce a new variable $\mathbf{Z}$. Then, Eq.~(11) is equivalent to
\begin{equation}
\begin{aligned}
  &\mathop{\text{minimize}} \limits_{\mathbf{X}} \quad \Vert \mathbf{X} \Vert_{*} +(1/2\lambda)\Vert \mathbf{B}\cdot \mathbf{Z}-\mathbf{B}\cdot \mathbf{M} \Vert_{F}^{2}\\
 & \text{subject to}  \qquad \qquad \mathbf{X}-\mathbf{Z}=0 .
\end{aligned}
\end{equation}

The augmented Lagrangian of (12) becomes
\begin{equation}
\begin{aligned}
 \mathcal{L}_{\rho}(\mathbf{X},\mathbf{Z},\mathbf{Y})&= \Vert \mathbf{X}\Vert_{*}+(1/2\lambda)\Vert \mathbf{B}\cdot \mathbf{Z}-\mathbf{B} \cdot \mathbf{M}\Vert_{F}^{2}\\
 &+\mathbf{Y}^{T}(\mathbf{X}-\mathbf{Z})+(\rho/2)\Vert \mathbf{X}-\mathbf{Z}\Vert_{F}^{2}.
\end{aligned}
\end{equation}

For convenience, by combining the linear and quadratic terms in the augmented Lagrangian and scaling the dual variable, Eq.~(13) can be simplified as
\begin{equation}
\begin{aligned}
 \mathcal{L}_{\rho}(\mathbf{X},\mathbf{Z},\mathbf{Y})&= \Vert \mathbf{X}\Vert_{*}+(1/2\lambda)\Vert \mathbf{B}\cdot \mathbf{Z}-\mathbf{B} \cdot \mathbf{M}\Vert_{F}^{2}\\
 &+(\rho/2)\Vert \mathbf{X}-\mathbf{Z}+\mathbf{U}\Vert_{F}^{2}+\emph{const},
\end{aligned}
\end{equation}
where \emph{const} represents a constant term, $\mathbf{U}$ is the scaled dual variable~\cite{30}.
Using the scaled dual variable, we can get the iterations of ADMM as follows.
\begin{equation}
\begin{aligned}
 \mathbf{X}^{k+1}&=\mathop{\text{argmin}} \limits_{\mathbf{X}} (\Vert \mathbf{X}\Vert_{*}+(\rho/2)\Vert \mathbf{X}-\mathbf{Z}^{k}+\mathbf{U}^{k}\Vert_{F}^{2}),\\
 \mathbf{Z}^{k+1}&=\mathop{\text{argmin}} \limits_{\mathbf{Z}}((1/2\lambda)\Vert \mathbf{B}\cdot \mathbf{Z}-\mathbf{B} \cdot \mathbf{M}\Vert_{F}^{2},\\
 &+(\rho/2)\Vert \mathbf{Z}-\mathbf{X}^{k+1}-\mathbf{U}^{k}\Vert_{F}^{2})\\
 \mathbf{U}^{k+1}&=\mathbf{U}^{k}+\mathbf{X}^{k+1}-\mathbf{Z}^{k+1}.
\end{aligned}
\end{equation}

\subsubsection{Update $\mathbf{X}$}
\

Before giving the update step for $\mathbf{X}$, we need the following definition and theorem.
\begin{defn}
  Assume that the singular value decomposition of matrix $\mathbf{X}$ is given by $\mathbf{X}=\mathbf{U}\text{diag}(\bm{\sigma})\mathbf{V}^{T}$, where \bm{$\sigma$} is the singular values vector of $\mathbf{X}$. $(\cdot)^{T}$ is the transpose operator, and $\mathbf{U}$ and $\mathbf{V}$ are orthogonal matrices. For any $\tau >0$, the matrix shrinkage operator $\mathscr{D}_{\tau}(\cdot)$ is defined as \rm{\cite{31}}.
  \begin{equation}
    \mathscr{D}_{\tau}(\mathbf{X})=\mathbf{U}\bm{\Sigma}_{\tau}\mathbf{V}^{T},
  \end{equation}
  where $\bm{\Sigma}_{\tau}=\text{diag}(\text{max}(\bm{\sigma}-\tau,0))$.
\end{defn}

\begin{thm}
   For any $\tau>0$, $\mathbf{X}:=\mathscr{D}_{\tau}(\mathbf{Y})$ is a closed form solution for the following optimization problem \rm{\cite{31}}.
  \begin{equation}
   \begin{aligned}
     \mathop{\text{minimize}} \limits_{\mathbf{X}} \quad f(\mathbf{X})=\tau\Vert \mathbf{X} \Vert_{*} +(1/2)\Vert \mathbf{X}-\mathbf{Y}\Vert_{F}^{2}.
    \end{aligned}
  \end{equation}
\end{thm}
The proof of Theorem 1 is detailed in~\cite{31}.

According to Eq.~(15), Definition 2, and Theorem 1, we can obtain the update of X.
\begin{equation}
 \mathbf{X}^{k+1}=\mathscr{D}_{1/\rho}(\mathbf{U}^{k}-\mathbf{Z}^{k}).
\end{equation}

\subsubsection{Update $\mathbf{Z}$}
\

Before giving the update step for $\mathbf{Z}$, we give the following proposition.
\begin{prop}
 Let $\mathbf{Z}_{left} = (1/\lambda)\mathbf{B}+\rho \mathbf{I}$, $\mathbf{Z}_{right}=(1/\lambda)(\mathbf{B}\cdot \mathbf{M})+\rho (\mathbf{X}+\mathbf{U})$. Then $\mathbf{Z}:=\mathbf{Z}_{right}./\mathbf{Z}_{left}$ is the closed form solution for the following optimization problem.
 \begin{equation}
\mathop{\text{minimize}}\limits_{\mathbf{Z}} \,(1/2\lambda)\Vert \mathbf{B}\cdot \mathbf{Z}-\mathbf{B} \cdot \mathbf{M}\Vert_{F}^{2}
 +(\rho/2)\Vert \mathbf{X}-\mathbf{Z}+\mathbf{U}\Vert_{F}^{2},
\end{equation}
where $(./)$ represents element-wise division of matrix and $\mathbf{I}$ is the identity matrix whose entries are all ones.
\end{prop}

 \textbf{Proof}: Suppose $\mathbf{Z}^{*}$ is the optimal solution to Eq.~(19), if and only if
\begin{equation}
 0 = (1/\lambda)(\mathbf{B}\cdot \mathbf{Z}^{k+1}-\mathbf{B}\cdot \mathbf{M})+\rho(\mathbf{Z}^{k+1}-\mathbf{X}^{k+1}-\mathbf{U}^{k}),
\end{equation}
which is equivalent to
\begin{equation}
  ((1/\lambda)\mathbf{B}+\rho \mathbf{I})\cdot \mathbf{Z}^{k+1}=(1/\lambda)(\mathbf{B}\cdot \mathbf{M})+\rho (\mathbf{X}^{k+1}+\mathbf{U}^{k}).
\end{equation}
Thus $\mathbf{Z}^{*} = \mathbf{Z}_{right}./\mathbf{Z}_{left}$.\QEDA

According to Eq.~(15) and Proposition 1, the update for variable $\mathbf{Z}$ becomes
 \begin{equation}
  \mathbf{Z}^{k+1}=\mathbf{Z}_{right}./\mathbf{Z}_{left}.
\end{equation}

\subsubsection{The ADRM Algorithm}
\

After discussing the appearing subproblem, we now present the complete ADRM algorithm for reconstruction of single-attribute sensor-data.

The algorithm inputs the sampling binary index matrix $\mathbf{B}$, incomplete sensor data matrix $\mathbf{M}$, and the parameters $\lambda,\rho,c_{\lambda},\lambda^{*}$. It iteratively minimizes (8) by decreasing $\lambda$ towards convergence. $\lambda^{*}$ is set to be the lower bound of parameter $\lambda$. Alg.~1 details the ADRM algorithm.
\begin{algorithm}[!htb]
\caption{ADRM algorithm for reconstruction of single-attribute sensor data}
\begin{algorithmic}[1]
  \STATE Given $\mathbf{B},\mathbf{M},\lambda,\rho,c_{\lambda},\lambda^{*}$
  \STATE Initialize $\mathbf{Z}^{0}=\mathbf{U}^{0}=0,k=0$
 \FOR{$k=0,1,...$}
   \STATE $\mathbf{X}^{k+1}=\mathscr{D}_{1/\rho}(\mathbf{U}^{k}-\mathbf{Z}^{k})$\\
   \STATE Calculate\; $\mathbf{Z}_{left} = (1/\lambda)\mathbf{B}+\rho \mathbf{I}$\\
   \STATE   \qquad \qquad \;  $\mathbf{Z}_{right}=(1/\lambda)(\mathbf{B}\cdot \mathbf{M})+\rho (\mathbf{X}^{k+1}+\mathbf{U}^{k})$\\
   \STATE $\mathbf{Z}^{k+1}=\mathbf{Z}_{right}./\mathbf{Z}_{left}$\\
   \STATE $\mathbf{U}^{k+1}=\mathbf{U}^{k}+\mathbf{X}^{k+1}-\mathbf{Z}^{k+1}$\\
   \STATE $\lambda ^{k+1}=\text{max}(c_{\lambda}\lambda^{k},\lambda^{*})$\\
 \ENDFOR
  \RETURN $\mathbf{X}^{k}$
  \end{algorithmic}
  \end{algorithm}

\section{Reconstruction of Multi-Attribute Sensor Data}
Tensor is the higher-order generalization of vector and matrix. It may better represent practical data structures. For example, sensor nodes in Internet of things can sense multiple-attribute data simultaneously, e.g., node in data sensing lab~\cite{32} senses temperature, humidity and microphone. Using tensor-based model to represent multiple-attribute sensor data can take full advantage of the correlations between attributes. It may further improve the accuracy of data reconstruction. In this section, we will follow the tensor completion method to solve the multi-attribute sensor-data reconstruction problem.

\subsection{Notation for Tensor}
We follow paper~\cite{34} to denote tensors with calligraphic font (e.g., $\mathcal{X}$). An $N$-order tensor is defined as $\mathcal{X} \in \mathbb{R}^{I_{1}\times I_{2}\times ...\times I_{N}}$. The ``unfold'' operation along the $k^{th}$ mode on a tensor $\mathcal{X}$ is defined as $\text{unfold}_{k}(\mathcal{X}):=\mathcal{X}_{(k)}\in \mathbb{R}^{I_{k}\times (I_{1}...I_{k-1}I_{k+1}...I_{N})}$. The opposite operation ``fold'' is defined as $\text{fold}_{k}(\mathcal{X}_{(k)}):=\mathcal{X}$.
$\Vert \mathcal{X}\Vert_{F}$ is the Frobenius norm of tensor.
The inner product of two identical-sized tensors $\mathcal{X},\mathcal{Y}\in \mathbb{R}^{I_{1}\times I_{2}\times ...\times I_{N}}$ is the sum of the products of their entries, i.e.,
\begin{equation}
<\mathcal{X},\mathcal{Y}>=\sum_{i_{1}=1}^{n_{1}}\sum_{i_{2}=1}^{n_{2}}\cdot\cdot\cdot\sum_{i_{N}=1}^{n_{N}} x_{i_{1}i_{2}...i_{N}}y_{i_{1}i_{2}...i_{N}}.
\end{equation}

\subsection{Tensor Low n-Rank Minimization Based Approach}
Assume $\mathcal{T}$ is the constructed $n$-order tensor sensor data, $\mathcal{T} \in \mathbb{R}^{I_{1}\times I_{2}\times ...\times I_{N}}$. Due to low-rank structure feature, the missing data in $\mathcal{T}$ can also be recovered by rank minimization. Generalize the matrix rank minimization model to higher-order tensor by solving the following optimization problem.
\begin{equation}
\begin{aligned}
&\mathop{\text{minimize}} \limits_{\mathcal{X}}   \qquad \quad \quad  \Vert \mathcal{X}\Vert _{*}\\
& \text{subject to} \quad \mathscr{P}_{\Omega}(\mathcal{X})=\mathscr{P}_{\Omega}(\mathcal{T}),
\end{aligned}
\end{equation}
where tensor nuclear norm is defined in~\cite{25},
\begin{equation}
 \Vert \mathcal{X}\Vert _{*}: =\sum \limits_{i=1}^{N} \alpha_{i}\Vert \mathcal{X}_{(i)}\Vert_{*},
\end{equation}
where $\alpha_{i}$ can be regarded as a weight to $\mathcal{X}_{(i)}$. Without loss of generality, here we let $\alpha_{i}=\alpha_{i+1}, i=1,2,...,N$, which means each $\mathcal{X}_{(i)}$ gets equal importance. Following this definition, the optimization in Eq.~(24) can be written as
\begin{equation}
 \begin{aligned}
  &\mathop{\text{minimize}} \limits_{\mathcal{X}} \qquad  \sum \limits_{i=1}^{N}  \Vert \mathcal{X}_{(i)}\Vert _{*}\\
  & \text{subject to} \quad \mathscr{P}_{\Omega}(\mathcal{X})=\mathscr{P}_{\Omega}(\mathcal{T}).
  \end{aligned}
\end{equation}

Consider that in practice, noises in sensory data may lead to over-fitting problem if strict satisfaction is required. Thus, we consider the following unconstrained problem.
\begin{equation}
\mathop{\text{minimize}} \limits_{\mathcal{X}} \quad \sum \limits_{i=1}^{N}  \Vert \mathcal{X}_{(i)}\Vert _{*}+ (1/2\lambda)\Vert \mathscr{P}_{\Omega}(\mathcal{X})-\mathscr{P}_{\Omega}(\mathcal{T})\Vert_{F}^{2},
\end{equation}
where parameter $0<\lambda \le 1$. The parameter $\lambda$ controls the fit to constraint $\mathscr{P}_{\Omega}(\mathcal{X})=\mathscr{P}_{\Omega}(\mathcal{T})$. Considering a continuation technique for decreasing the value of $\lambda$ towards convergence, the $\mathscr{P}_{\Omega}(\mathcal{X})$ is close to but not equal to $\mathscr{P}_{\Omega}(\mathcal{T})$.

Likewise, for convenience, we define sampling tensor $\mathcal{B}$,
\begin{equation}
b_{ij..n} = \left\{
     \begin{array}{lcl}
     {1, \quad \text{if}\;(i,j,..,n) \in \Omega}\\
     {0, \quad \text{otherwise}},
     \end{array}
     \right.
\end{equation}
where $\Omega$ is the observing data set.

Now, we define the multi-attribute sensor-data reconstruction problem.
\begin{defn}
  Let $\mathcal{B}$ denote binary sampling tensor and $\mathcal{T}$ denote incomplete tensor of multi-attribute sensor data. Then the missing values in $\mathcal{T}$ can be effectively estimated by solving the convex optimization problem below,
\begin{equation}
\mathop{\text{minimize}} \limits_{\mathcal{X}} \quad \sum \limits_{i=1}^{N}  \Vert \mathcal{X}_{(i)}\Vert _{*}+ (1/2\lambda)\Vert\mathcal{B} \cdot \mathcal{X}-\mathcal{B} \cdot \mathcal{T}\Vert_{F}^{2},
\end{equation}
  where ($\cdot$) denotes the element-wise production of tensor, $\lambda$ is the parameter.
\end{defn}
In the following part of this section, we propose an ADMM based multi-attribute sensor-data reconstruction algorithm, namely ADMAR, to solve Eq.~(29).

\subsection{ADMAC Algorithm for Reconstruction of Multi-Attribute Sensor Data}
In order to apply the ADMM method to Eq.~(29), we need to transform it into ADMM  form. Thus we need to perform variable splitting.

We introduce $N$ new tensor-valued variables, $\mathcal{Y}_{1},...,\mathcal{Y}_{N}$. Let $\mathcal{Y}_{i}=\mathcal{X}, i \in \{1,...,N\}$. With these new variables $\mathcal{Y}_{i}$, Eq.~(29) can be rephrased as follows.
\begin{equation}
\begin{aligned}
&\mathop{\text{minimize}} \limits_{\mathcal{X},\mathcal{Y}} \; \sum \limits_{i=1}^{N} \Vert \mathcal{Y}_{i,(i)}\Vert_{*}+
(1/2\lambda)\Vert \mathcal{B} \cdot \mathcal{X}-\mathcal{B} \cdot \mathcal{T}\Vert_{F}^{2}\\
&\text{subject to} \qquad \mathcal{Y}_{i}=\mathcal{X}, \quad i=1,...,N.
\end{aligned}
\end{equation}

The augmented Lagrangian of Eq.~(30) is
\begin{equation}
\begin{aligned}
\mathcal{L}_{\rho}(\mathcal{Y}_{i},\mathcal{X},\mathcal{U}_{i})&=\sum \limits_{i=1}^{N} \Vert \mathcal{Y}_{i,(i)}\Vert_{*}+
(1/2\lambda)\Vert \mathcal{B} \cdot \mathcal{X}-\mathcal{B} \cdot \mathcal{T}\Vert_{F}^{2}\\
&+(\rho/2)\sum \limits_{i=1}^{N}\Vert \mathcal{Y}_{i}-\mathcal{X}+\mathcal{U}_{i}\Vert_{F}^{2},
\end{aligned}
\end{equation}
where $\mathcal{U}_{i},i=1,...,N$ is the scaled dual variable.
We can get the iterations of ADMM.
\begin{equation}
\begin{aligned}
\mathcal{Y}_{i}^{k+1}&=\mathop{\text{argmin}} \limits_{\mathcal{Y}_{i}}(\Vert \mathcal{Y}_{i,(i)}\Vert_{*}+(\rho/2)\Vert \mathcal{Y}_{i}-\mathcal{X}^{k}+\mathcal{U}_{i}^{k}\Vert_{F}^{2}),\\
\mathcal{X}^{k+1}&=\mathop{\text{argmin}} \limits_{\mathcal{X}}((1/2\lambda)\Vert
\mathcal{B} \cdot \mathcal{X}-\mathcal{B} \cdot \mathcal{T}\Vert_{F}^{2}\\
 &+(\rho/2)\sum \limits_{i=1}^{N}\Vert \mathcal{X}-\mathcal{Y}_{i}^{k+1}-\mathcal{U}_{i}^{k}\Vert_{F}^{2}),\\
\mathcal{U}_{i}^{k+1}&=\mathcal{U}_{i}^{k}+\mathcal{Y}_{i}^{k+1}-\mathcal{X}^{k+1}.
\end{aligned}
\end{equation}

\subsubsection{Update $\mathcal{Y}$}
\

Variable $\mathcal{Y}_{i}$ can be solved independently by the matrix shrinkage operator introduced in Section III.
So the update for $\mathcal{Y}_{i}$ becomes
\begin{equation}
 \mathcal{Y}_{i}^{k+1}=\text{fold}_{i}(\mathscr{D}_{1/\rho}(\mathcal{X}^{k}-\mathcal{U}_{i}^{k})_{(i)}).
\end{equation}

\subsubsection{Update $\mathcal{X}$}
\

Before giving the update for $\mathcal{X}$, we provide the following proposition.
\begin{prop}
  Let $\overline{\mathcal{Y}}=(1/N)\sum \limits_{i=1}^{N}\mathcal{Y}_{i}$, $\overline{\mathcal{U}}=(1/N)\sum \limits_{i=1}^{N}\mathcal{U}_{i}$, $\mathcal{X}_{left}=(1/\lambda)\mathcal{B}+N\rho \mathcal{I},\;  \mathcal{X}_{right}=(1/\lambda)(\mathcal{B}\cdot \mathcal{T})+N\rho (\overline{\mathcal{Y}}+\overline{\mathcal{U}})$. Then $\mathcal{X}:=\mathcal{X}_{right}./\mathcal{X}_{left}$ is the closed form solution to
  \begin{equation}
  \mathop{\text{min}} \limits_{\mathcal{X}} \,(1/2\lambda)\Vert
  \mathcal{B}\cdot \mathcal{X}-\mathcal{B}\cdot \mathcal{T}\Vert_{F}^{2}+(\rho/2)\sum \limits_{i=1}^{N}\Vert \mathcal{X}-\mathcal{Y}_{i}-\mathcal{U}_{i}\Vert_{F}^{2}
  \end{equation}
where $\mathcal{I}$ is a tensor with all its entries set to one and $(\cdot /)$ denotes the element-wise division of tensor.
\end{prop}
\textbf{Proof}: Suppose $\mathcal{X}^{*}$ is the optimal solution to Eq.~(34), if and only if
\begin{equation}
 0=(1/\lambda)(\mathcal{B}\cdot \mathcal{X}^{*}-\mathcal{B}\cdot \mathcal{T})+\rho(N\mathcal{X}^{*}-\sum \limits_{i=1}^{N}(\mathcal{Y}_{i}+\mathcal{U}_{i})).
 \end{equation}
Substitute $\overline{\mathcal{Y}}$ and $\overline{\mathcal{U}}$ into the above equation, then Eq.~(35) is equivalent to
\begin{equation}
 ((1/\lambda)\mathcal{B}+N\rho \mathcal{I})\cdot \mathcal{X}^{*}=(1/\lambda)(\mathcal{B}\cdot \mathcal{T})+N\rho (\overline{\mathcal{Y}}+\overline{\mathcal{U}}).
\end{equation}
Thus $\mathcal{X}^{*}=\mathcal{X}_{right}./\mathcal{X}_{left}$. \QEDA

According to Eq.~(32) and Proposition 2, we obtain the update for $\mathcal{X}^{k+1}$£º
\begin{equation}
 \mathcal{X}^{k+1}=\mathcal{X}_{right}./\mathcal{X}_{left}.
\end{equation}

\subsubsection{The ADMAC Algorithm}
\

After discussing the appearing subproblem, we present the complete ADMAC algorithm for multi-attribute sensor-data reconstruction, shown in Alg. 2.

The algorithm inputs the binary sampling tensor $\mathcal{B}$, incomplete sensor data tensor $\mathcal{T}$, and the parameters $\lambda,\rho,c_{\lambda},\lambda^{*}$. It iteratively minimizes Eq.~(29) by decreasing $\lambda$ towards convergence. $\lambda^{*}$ is set to be the lower bound of parameter $\lambda$.
\begin{algorithm}[!htb]
\caption{ADMAC algorithm for multi-attribute sensor-data reconstruction }
\begin{algorithmic}[1]
  \STATE Given $\mathcal{B},\mathcal{T},\lambda,\rho,c_{\lambda},\lambda^{*}$
  \STATE Initialize $\mathcal{X}^{0}=\mathcal{U}_{i}^{0}=0, i=1,...,N, k=0$
  \FOR{$k=0,1,...$}
    \FOR{$i=1:N$}
      \STATE $\mathcal{Y}_{i}^{k+1}=\text{fold}_{i}(\mathscr{D}_{1/\rho}(\mathcal{X}^{k}-\mathcal{U}_{i}^{k})_{(i)})$
    \ENDFOR
    \STATE Calculate $\overline{\mathcal{Y}}^{k+1}=(1/N)\sum \limits_{i=1}^{N}\mathcal{Y}_{i}^{k+1}$
    \STATE \qquad \qquad \;$\overline{\mathcal{U}}^{k}=(1/N)\sum \limits_{i=1}^{N}\mathcal{U}_{i}^{k}$
    \STATE \qquad \qquad  $\mathcal{X}_{left}=(1/\lambda)\mathcal{B}+N\rho \mathcal{I}$
    \STATE \qquad \qquad  $\mathcal{X}_{right}=(1/\lambda)(\mathcal{B}\cdot \mathcal{T})+N\rho(\overline{\mathcal{Y}}^{k+1}+\overline{\mathcal{U}}^{k})$
    \STATE $\mathcal{X}^{k+1}=\mathcal{X}_{right}./\mathcal{X}_{left}$
    \FOR{$i=1:N$}
      \STATE $\mathcal{U}_{i}^{k+1}=\mathcal{U}_{i}^{k}+\mathcal{Y}_{i}^{k+1}-\mathcal{X}^{k+1}$
    \ENDFOR
    \STATE $\lambda ^{k+1}=\text{max}(c_{\lambda}\lambda^{k},\lambda^{*})$
  \ENDFOR
  \RETURN $\mathcal{X}^{k}$
  \end{algorithmic}
  \end{algorithm}

\subsection{The HaLRTC Algorithm}
In this subsection, we briefly introduce the HaLRTC algorithm, which is proposed in~\cite{25} to estimate the missing values in visual data without observation noise. In this paper, however, we use it to reconstruct missing sensor data and compare its performance with our proposed ADMAC algorithm.

Instead of using relaxation technique to relax Eq.~(26) into unconstrained formulation, HaLRTC algorithm handles this equality directly. By introducing $N$ new tensor-valued variables, $\mathcal{Y}_{1},...,\mathcal{Y}_{N}$, and let $\mathcal{Y}_{i,(i)}=\mathcal{X}_{(i)}, i \in \{1,...,N\}$, then Eq.~(26) becomes
\begin{equation}
  \begin{aligned}
  &\mathop{\text{minimize}} \limits_{\mathcal{X}} \quad \sum \limits_{i=1}^{N}  \Vert \mathcal{Y}_{i,(i)}\Vert _{*}\\
  & \text{subject to} \quad
   \mathscr{P}_{\Omega}(\mathcal{X})=\mathscr{P}_{\Omega}(\mathcal{T})\\
  & \qquad \quad \quad \quad \mathcal{Y}_{i}=\mathcal{X},i=1,...,N.
   \end{aligned}
\end{equation}
The augmented Lagrangian function is as follows.
\begin{equation}
 \mathscr{L}_{\rho}(\mathcal{X},\mathcal{Y}_{i},\mathcal{U}_{i})=\sum \limits_{i=1}^{N} (\Vert \mathcal{Y}_{i,{i}}\Vert_{*}+(\rho/2)\Vert \mathcal{Y}_{i}-\mathcal{X}+\mathcal{U}_{i}\Vert_{F}^{2}),
\end{equation}
where $\mathcal{U}_{i}, i=1,...,N$ is the scaled dual variable. According to the framework of ADMM, $\mathcal{Y}_{i},\mathcal{X},\mathcal{U}_{i}$ can be iteratively updated. The HaLRTC algorithm is listed in Alg.~3.
\begin{algorithm}[!htb]
\caption{HaLRTC algorithm for multi-attribute sensor-data reconstruction }
\begin{algorithmic}[1]
  \STATE Given $\mathcal{B},\mathcal{T},\rho$
  \STATE Initialize $\mathcal{U}_{i}^{0}=0, i=1,...,N, k=0$
  \STATE Set $\mathscr{P}_{\Omega}(\mathcal{X})=\mathscr{P}_{\Omega}(\mathcal{T})$ and $\mathscr{P}_{\overline{\Omega}}(\mathcal{X}^{0})=0, where \overline{\Omega}$ denotes the complementary set of $\Omega$, $\mathcal{X}^{0}=\mathscr{P}_{\Omega}(\mathcal{X})+\mathscr{P}_{\overline{\Omega}}(\mathcal{X}^{0})$
  \FOR{$k=0,1,...$}
   \FOR{$i=1:N$}
    \STATE $\mathcal{Y}_{i}^{k+1}=\text{fold}_{i}(\mathscr{D}_{1/\rho}(\mathcal{X}^{k}-\mathcal{U}_{i}^{k})_{(i)})$
   \ENDFOR
   \STATE $\mathscr{P}_{\overline{\Omega}}(\mathcal{X}^{k+1})=(1/N)\mathscr{P}_{\overline{\Omega}}(\sum \limits_{i=1}^{N}(\mathcal{Y}_{i}^{k+1}+\mathcal{U}_{i}^{k}))$
   \STATE $\mathcal{X}^{k+1}=\mathscr{P}_{\Omega}(\mathcal{X})+\mathscr{P}_{\overline{\Omega}}(\mathcal{X}^{k+1})$
   \FOR{$i=1:N$}
     \STATE $\mathcal{U}_{i}^{k+1}=\mathcal{U}_{i}^{k}+\mathcal{Y}_{i}^{k+1}-\mathcal{X}^{k+1}$
   \ENDFOR
  \ENDFOR
  \RETURN $\mathcal{X}^{k}$
  \end{algorithmic}
  \end{algorithm}

\section{Relaxed Version of Multi-Attribute Sensor-Data Reconstruction}
In Section IV, we assume that the constructed tensor using multiple-attribute sensor data is jointly low-rank in all modes, which might be too strict to be satisfied in practice. The mixture model for a low-rank tensor is introduced in~\cite{35}, which only requires the tensor to be the sum of a set of component tensors, and each of which is low-rank in the corresponding mode, i.e., $\mathcal{X}=\sum \limits_{i=1}^{N} \mathcal{X}_{i}$, where $\mathcal{X}_{i,(i)}$ is a low-rank matrix for each $i$-mode. It is shown in~\cite{35} that the mixture model can automatically detect the rank-deficient mode and yield better recovery performance when the original tensor is low-rank only in certain modes.

Based on this mixture model, we modify Definition 3 and redefine the multi-attribute sensor-data reconstruction problem.
\begin{defn}
  Let $\mathcal{B}$ denote binary sampling tensor and $\mathcal{T}$ denote the incomplete tensor of multi-attribute sensor data. Then the missing values in $\mathcal{T}$ can be effectively estimated by solving the convex optimization problem below.
\begin{equation}
\mathop{\text{minimize}} \limits_{\mathcal{X}} \ \sum \limits_{i=1}^{N}  \Vert \mathcal{X}_{i,(i)}\Vert _{*}+ (1/2\lambda)\Vert\mathcal{B} \cdot \sum \limits_{i=1}^{N} \mathcal{X}_{i}-\mathcal{B} \cdot \mathcal{T}\Vert_{F}^{2},
\end{equation}
  where ($\cdot$) denotes the element-wise production of tensor.
\end{defn}

In the following part of this section, we utilize the ADMM method to solve Eq.~(40) and propose an algorithm named relaxed version of ADMM based multi-attribute sensor-data completion algorithm, namely, R-ADMAC.

\subsection{R-ADMAC Algorithm for Multi-Attribute Sensor-Data Reconstruction}
Likewise, we first transform Eq.~(40) into ADMM form by introducing $N$ new tensor-valued variables $\mathcal{Z}_{1},...,\mathcal{Z}_{N}$. Let $\mathcal{X}_{i}=\mathcal{Z}_{i}, i\in \{1,...,N\}$. With these new variables $\mathcal{Z}_{i}$, Eq.~(40) can be rewritten as follows.
\begin{equation}
\begin{aligned}
&\mathop{\text{minimize}} \limits_{\mathcal{X}_{i},\mathcal{Z}_{i}}  \sum \limits_{i=1}^{N} \Vert \mathcal{X}_{i,(i)}\Vert_{*}+
(1/2\lambda)\Vert \mathcal{B} \cdot \sum \limits_{i=1}^{N} \mathcal{Z}_{i}-\mathcal{B} \cdot \mathcal{T}\Vert_{F}^{2}\\
&\text{subject to} \qquad \mathcal{X}_{i}-\mathcal{Z}_{i}, \quad i=1,...,N.
\end{aligned}
\end{equation}

The augmented Lagrangian of Eq.~(41) is
\begin{equation}
\begin{aligned}
\mathcal{L}_{\rho}(\mathcal{X}_{i},\mathcal{Z}_{i},\mathcal{U}_{i})&=\sum \limits_{i=1}^{N} \Vert \mathcal{X}_{i,(i)}\Vert_{*}+
(1/2\lambda)\Vert \mathcal{B} \cdot \sum \limits_{i=1}^{N} \mathcal{Z}_{i}-\mathcal{B} \cdot \mathcal{T}\Vert_{F}^{2}\\
&+(\rho/2)\sum \limits_{i=1}^{N}\Vert \mathcal{X}_{i}-\mathcal{Z}_{i}+\mathcal{U}_{i}\Vert_{F}^{2},
\end{aligned}
\end{equation}
where $\mathcal{U}_{i},i=1,...,N$, is the scaled dual variable. Let $f_{i}(\mathcal{X}_{i})=\Vert \mathcal{X}_{i,(i)}\Vert_{*}$, $g(\sum \limits_{i=1}^{N} \mathcal{Z}_{i})=(1/2\lambda)\Vert \mathcal{B} \cdot \sum \limits_{i=1}^{N} \mathcal{Z}_{i}-\mathcal{B} \cdot \mathcal{T}\Vert_{F}^{2}$. We can get the iterations of ADMM.
\begin{equation}
\begin{aligned}
\mathcal{X}_{i}^{k+1}&=\mathop{\text{argmin}} \limits_{\mathcal{X}_{i}}( f_{i}(\mathcal{X}_{i})+(\rho/2)\Vert \mathcal{X}_{i}-\mathcal{Z}_{i}^{k}+\mathcal{U}_{i}^{k}\Vert_{F}^{2}),\\
\mathcal{Z}^{k+1}&=\mathop{\text{argmin}} \limits_{\mathcal{Z}_{i}}( g(\sum \limits_{i=1}^{N} \mathcal{Z}_{i})+(\rho/2)\sum \limits_{i=1}^{N}\Vert \mathcal{Z}_{i}-\mathcal{X}_{i}^{k+1}-\mathcal{U}_{i}^{k}\Vert_{F}^{2}),\\
\mathcal{U}_{i}^{k+1}&=\mathcal{U}_{i}^{k}+\mathcal{X}_{i}^{k+1}-\mathcal{Z}_{i}^{k+1}.
\end{aligned}
\end{equation}

From the iteration equations introduced in Eq.~(43), it can be seen that $\mathcal{X}_{i}$ and $\mathcal{U}_{i}$ can be carried out independently in parallel for each $i=1,...,N$. However, it is tricky for $\mathcal{Z}$-update. In the following part, we will provide a method to solve this problem.
\begin{prop}
 The $\mathcal{Z}$-update is equal to solving the following unconstrained problem,
 \begin{equation}
  \mathop{\text{minimize}} \limits_{\mathcal{\overline{Z}}} \ g(N\mathcal{\overline{Z}}) + (\rho/2)\sum \limits_{i=1}^{N}\Vert \mathcal{\overline{Z}}-\mathcal{\overline{X}}^{k+1}- \mathcal{\overline{U}}^{k} \Vert_{F}^{2}.
\end{equation}
where $\mathcal{\overline{Z}}=(1/N)\sum \limits_{i=1}^{N} \mathcal{Z}_{i}, \mathcal{\overline{X}}^{k+1}=(1/N)\sum \limits_{i=1}^{N} \mathcal{X}_{i}^{k+1},\mathcal{\overline{U}}^{k}=(1/N)\sum \limits_{i=1}^{N} \mathcal{U}_{i}^{k}$.
\end{prop}

\textbf{Proof}: In order to simplify notations, let $\mathcal{P}_{i}=\mathcal{X}_{i}^{k+1}+\mathcal{U}_{i}^{k}$. Then, based on Eq. (43) the $\mathcal{Z}$-update can be rewritten as
\begin{equation}
\begin{aligned}
&\mathop{\text{minimize}} \quad g(N\mathcal{\overline{Z}})+(\rho/2)\sum \limits_{i=1}^{N}\Vert \mathcal{Z}_{i}-\mathcal{P}_{i}\Vert_{F}^{2}\\
&\text{subject to} \quad \qquad \mathcal{\overline{Z}}=(1/N)\sum \limits_{i=1}^{N} \mathcal{Z}_{i},
\end{aligned}
\end{equation}
By minimizing over $\mathcal{Z}_{1},...,\mathcal{Z}_{N}$ with additional variable $\mathcal{\overline{Z}}$ and fixed $\mathcal{\overline{Z}}$, we get the following solution.
\begin{equation}
  \mathcal{Z}_{i} = \mathcal{P}_{i}+\mathcal{\overline{Z}}-\mathcal{\overline{P}}.
\end{equation}

So the $\mathcal{Z}$-update can be computed by solving the unconstrained problem.
\begin{equation}
  \mathop{\text{minimize}} \limits_{\mathcal{\overline{Z}}} \quad g(N\mathcal{\overline{Z}}) + (\rho/2)\sum \limits_{i=1}^{N}\Vert \mathcal{\overline{Z}}-\mathcal{\overline{P}} \Vert_{F}^{2}.
\end{equation}
\QEDA

Then, substituting Eq. (46) for $\mathcal{Z}_{i}^{k+1}$ in the $\mathcal{U}$-update yields
\begin{equation}
  \mathcal{U}_{i}^{k+1} = \mathcal{\overline{U}}^{k}+\mathcal{\overline{X}}^{k+1}-\mathcal{\overline{Z}}^{k+1},
\end{equation}
which shows that the dual variables $\mathcal{U}_{i}^{k}$ are all equal and can be replaced by a single dual variable $\mathcal{U}$. By substituting for $\mathcal{Z}_{i}^{k}$ in the $\mathcal{X}$-update, the final iterations of ADMM becomes
\begin{equation}
\begin{aligned}
\mathcal{X}_{i}^{k+1}&=\mathop{\text{argmin}} \limits_{\mathcal{X}_{i}}( f_{i}(\mathcal{X}_{i})+(\rho/2)\Vert \mathcal{X}_{i}-\mathcal{Z}_{i}^{k}+\mathcal{\overline{X}}^{k}-\mathcal{\overline{Z}}^{k}+\mathcal{U}^{k}\Vert_{F}^{2}),\\
\mathcal{\overline{Z}}^{k+1}&=\mathop{\text{argmin}} \limits_{\mathcal{\overline{Z}}}( g(N\mathcal{\overline{Z}})+(N\rho/2)\Vert \mathcal{\overline{Z}}-\mathcal{\overline{X}}^{k+1}-\mathcal{U}^{k}\Vert_{F}^{2}),\\
\mathcal{U}^{k+1}&=\mathcal{U}^{k}+\mathcal{\overline{X}}^{k+1}-\mathcal{\overline{Z}}^{k+1}.
\end{aligned}
\end{equation}

\subsubsection{Update the $\mathcal{X}$-Variables}
\

Variable $\mathcal{X}_{i}$ can be solved independently by the matrix shrinkage operator introduced in Section III.
So the update for $\mathcal{Y}_{i}$ becomes
\begin{equation}
 \mathcal{X}_{i}^{k+1}=\text{fold}_{i}(\mathscr{D}_{1/\rho}(\mathcal{Z}_{i}^{k}-\mathcal{\overline{X}}^{k}+\mathcal{\overline{Z}}^{k}-\mathcal{U}^{k})_{(i)}).
\end{equation}

\subsubsection{Update the $\mathcal{\overline{Z}}$-Variable}
\

The $\mathcal{\overline{Z}}$-update is computed by solving the following equation
\begin{equation}
  \mathop{\text{minimize}} \quad g(N\mathcal{\overline{Z}})+(N\rho/2)\Vert \mathcal{\overline{Z}}-\mathcal{\overline{X}}^{k+1}-\mathcal{U}^{k}\Vert_{F}^{2}.
\end{equation}

Suppose $\mathcal{\overline{Z}}^{*}$ is the optimal solution of (51), if and only if
\begin{equation}
  0=(1/\lambda)(\mathcal{B} \cdot N\mathcal{\overline{Z}}^{*}-\mathcal{B} \cdot \mathcal{T})+N\rho(\mathcal{\overline{Z}}^{*}-\mathcal{\overline{X}}^{k+1}-\mathcal{U}^{k}),
\end{equation}
which is equivalent to
\begin{equation}
  ((1/\lambda)(\mathcal{B}+\rho\mathcal{I}) \cdot \mathcal{\overline{Z}}^{*} = \rho(\mathcal{\overline{X}}^{k+1}+\mathcal{U}^{k})+(1/(N\lambda))\mathcal{B} \cdot \mathcal{T}.
\end{equation}
 Let $\mathcal{Z}_{left}=(1/\lambda)\mathcal{B}+\rho \mathcal{I},\;  \mathcal{Z}_{right}=\rho(\mathcal{\overline{X}}^{k+1}+\mathcal{U}^{k})+(1/(N\lambda))\mathcal{B} \cdot \mathcal{T}$. We get the $\mathcal{\overline{Z}}$-update solution
 \begin{equation}
 \mathcal{\overline{Z}}^{k+1}=\mathcal{\overline{Z}}^{*}=\mathcal{Z}_{right}./\mathcal{Z}_{left}.
\end{equation}

\subsubsection{The R-ADMAC Algorithm}
\

After discussing the appearing subproblem, we present the complete R-ADMAC algorithm for multi-attribute sensor-data reconstruction, as shown in Alg.~4.

The algorithm inputs the sampling binary index tensor $\mathcal{B}$, incomplete sensor data tensor $\mathcal{T}$ and the parameters $\lambda,\rho,c_{\lambda},\lambda^{*}$. It iteratively minimizes Eq.~(40) by decreasing $\lambda$ toward convergence. $\lambda^{*}$ is set to be the lower bound of $\lambda$.

\begin{algorithm}[!htb]
\caption{R-ADMAC algorithm for multi-attribute sensor-data reconstruction }
\begin{algorithmic}[1]
  \STATE Given $\mathcal{B},\mathcal{T},\lambda,\rho,c_{\lambda},\lambda^{*}$
  \STATE Initialize $\mathcal{\overline{Z}}^{0}=\mathcal{U}^{0}=\mathcal{X}_{i}^{0}=0, i=1,...,N$
  \FOR{$k=0,1,...$}
    \FOR{$i=1:N$}
      \STATE $\mathcal{\overline{X}}^{k}=(1/N)\sum \limits_{i=1}^{N} \mathcal{X}_{i}^{k}$
      \STATE
      $\mathcal{X}_{i}^{k+1}=\text{fold}_{i}(\mathscr{D}_{1/\rho}(\mathcal{Z}_{i}^{k}-\mathcal{\overline{X}}^{k}+\mathcal{\overline{Z}}^{k}-\mathcal{U}^{k})_{(i)})$
    \ENDFOR
    \STATE Calculate $\mathcal{\overline{X}}^{k+1}=(1/N)\sum \limits_{i=1}^{N} \mathcal{X}_{i}^{k+1}$
    \STATE \qquad \qquad\;$\mathcal{Z}_{left}=(1/\lambda)\mathcal{B}+\rho \mathcal{I}$
    \STATE \qquad \qquad\;$\mathcal{Z}_{right}=\rho(\mathcal{\overline{X}}^{k+1}+\mathcal{U}^{k})+(1/(N\lambda))\mathcal{B} \cdot \mathcal{T}$
    \STATE $\mathcal{\overline{Z}}^{k+1}=\mathcal{Z}_{right}./\mathcal{Z}_{left}$

    \STATE$ \mathcal{U}^{k+1}=\mathcal{U}^{k}+\mathcal{\overline{X}}^{k+1}-\mathcal{\overline{Z}}^{k+1}$
    \STATE $\lambda ^{k+1}=\text{max}(c_{\lambda}\lambda^{k},\lambda^{*})$
  \ENDFOR
  \RETURN $\sum \limits_{i=1}^{N} \mathcal{X}_{i}^{k+1}$
\end{algorithmic}
\end{algorithm}

\section{Performance Evaluation}
In this section we evaluate the performance of the proposed algorithms and compare them with existing algorithms for missing data estimation in sensor data reconstruction.

\subsection{Experiment Setup}
We perform our study using two datasets, i.e., the ``Intel Berkeley" dataset and the ``Data Sensing Lab" dataset. For each dataset, we study two missing patterns of sensor data, i.e., ``random missing" and ``consecutive missing". In this section, we will describe these datasets and patterns, as well as the parameter settings for our proposed algorithms.

\subsubsection{Intel Berkeley Dataset}
\

 The data of Intel Berkeley experiment~\cite{33} are gathered by Intel Berkeley Research lab from February 28$^{th}$ to April 5$^{th}$, 2004. There are 54 Mica2Dot nodes placed in a 40m$\times$30m room. Every node reports once every 30 seconds. Sensor data include temperature, light, humidity, and voltage data.

\subsubsection{Data Sensing Lab Dataset}
\

 The data in the data sensing lab~\cite{32} are gathered by around 50 sensor motes distributed at the O'Reilly Strata Conference venue in Santa Clara in February, 2013. These motes are distributed around the conference venue and report back during the conference. Sensor data contain temperature, humidity, and microphone data.

\subsubsection{Missing Data}
\

Although both datasets have missing readings, we cannot directly use those for evaluation because their actual values are unknown. Instead, we first get complete raw sensor data from these two datasets, then produce artificial missing data with either random missing pattern or consecutive missing pattern.

The consecutive missing pattern means that some nodes miss all data after a certain sampling time point due to damage or running out of energy. In our consecutive missing pattern experiment, we randomly choose $10\%$ nodes as objective nodes occurring consecutive data missing and let each objective node miss last $x\%$ of all its data.

\subsubsection{Parameter Settings}
\

We employ error ratio to measure the differences between the predicted values and the actual values. The error ratio is a metric for measuring the reconstruction error and is defined as
\begin{equation}
 \epsilon = \frac{\sqrt{\sum_{(i,j)\in \overline{\Omega}}(x(i,j)-\hat{x}(i,j))^{2}}}{\sqrt{\sum_{(i,j)\in \overline{\Omega}}(x(i,j))^{2}}},
\end{equation}
where $\overline{\Omega}$ denotes the missing dataset.

In our experiment, sampling ratio means the observation ratio of sensor data, which is defined as
\begin{equation}
 \varepsilon = \frac{\sum_{(i,j)\in \Omega}1}{\sum_{(i,j)\in (\Omega \cup \overline{\Omega})}1},
\end{equation}
where $\Omega$ denotes the observation dataset.

As all the proposed algorithms are based on ADMM, we choose identical parameters for them. Specifically, $\lambda=1, c=1/4,\lambda^{*}=1e-6$. $\rho$ can be any positive number and its value affects the speed of convergence of the ADMM algorithm. In our experiment, we choose $\rho=0.1/std(\mathbf{y})$~\cite{35}, where $\mathbf{y}$ is a vector composed of all the observation values and $std(\mathbf{y})$ is the standard deviation of the observed values $\mathbf{y}$.

All the algorithms are implemented using MATLAB running on a desktop computer
with 3.2-GHz Intel i5-3470 CPU and 4 GB RAM. The experiments are repeated
30 times.

\subsection{Experiments of Single-Attribute Sensor-Data Reconstruction}
 In this section, we evaluate ADRM algorithm in the case of single-attribute sensor-data reconstruction. In order to verify the effectiveness of ADRM, we choose other two methods for comparison. One is the K-nearest neighbor (KNN) method. The other is the interior point method based on CVX~\cite{27}.

\subsubsection{Random Missing Pattern}
\

In this experiment, we calculate error ratios with different sampling ratios. The sampling ratios range from $10\%$ to $90\%$. Fig.~1 shows the experiment results, where X-axis represents sampling ratios, and Y-axis shows resulting error ratios. Generally speaking, error ratios decrease with sampling ratios.

\begin{figure}[!t]
\centering
\subfloat[Intel Berkeley (temperature)]{\includegraphics[width=4.4cm]{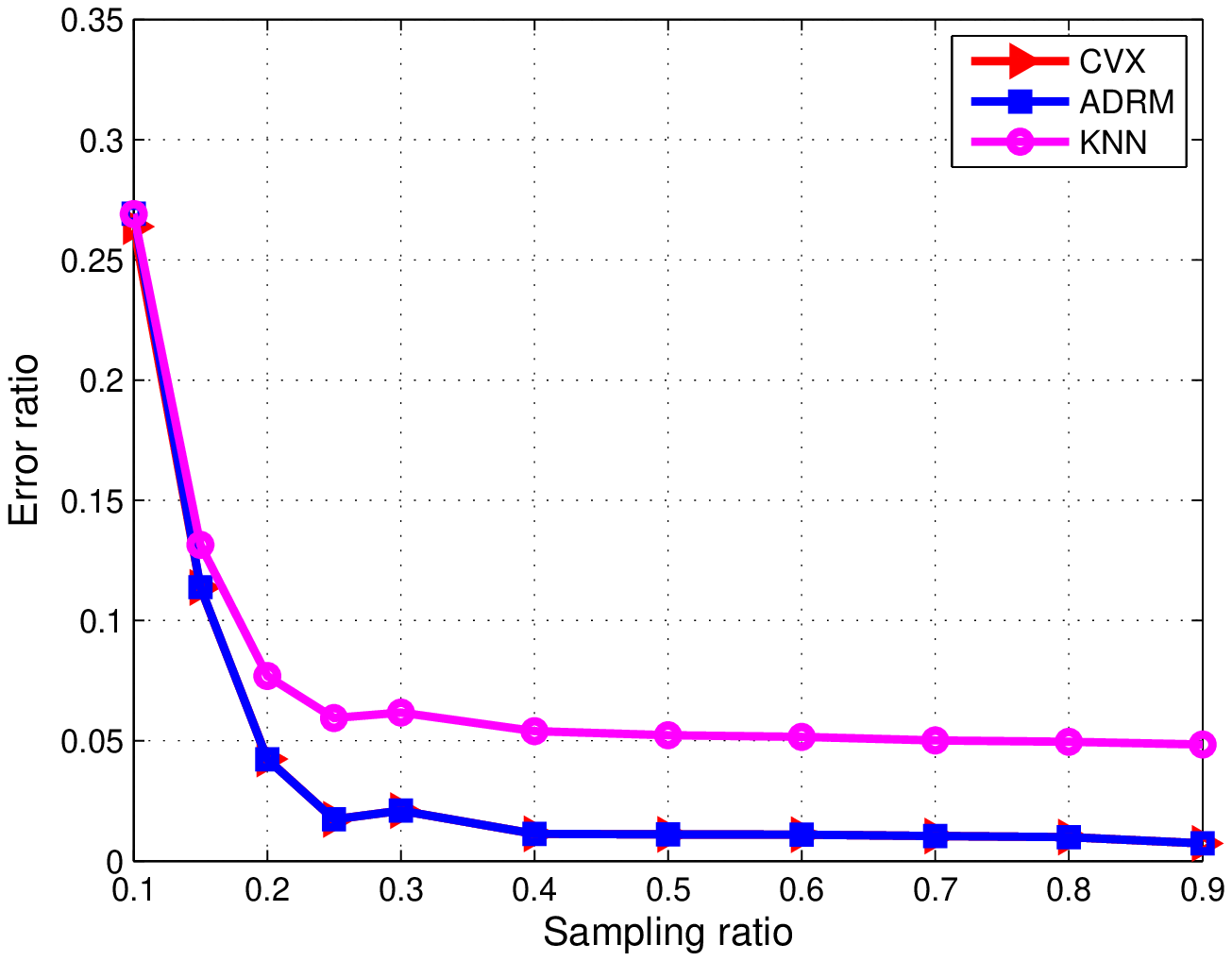}%
\label{fig1_1}}
\hfil
\subfloat[Intel Berkeley (humidity)]{\includegraphics[width=4.4cm]{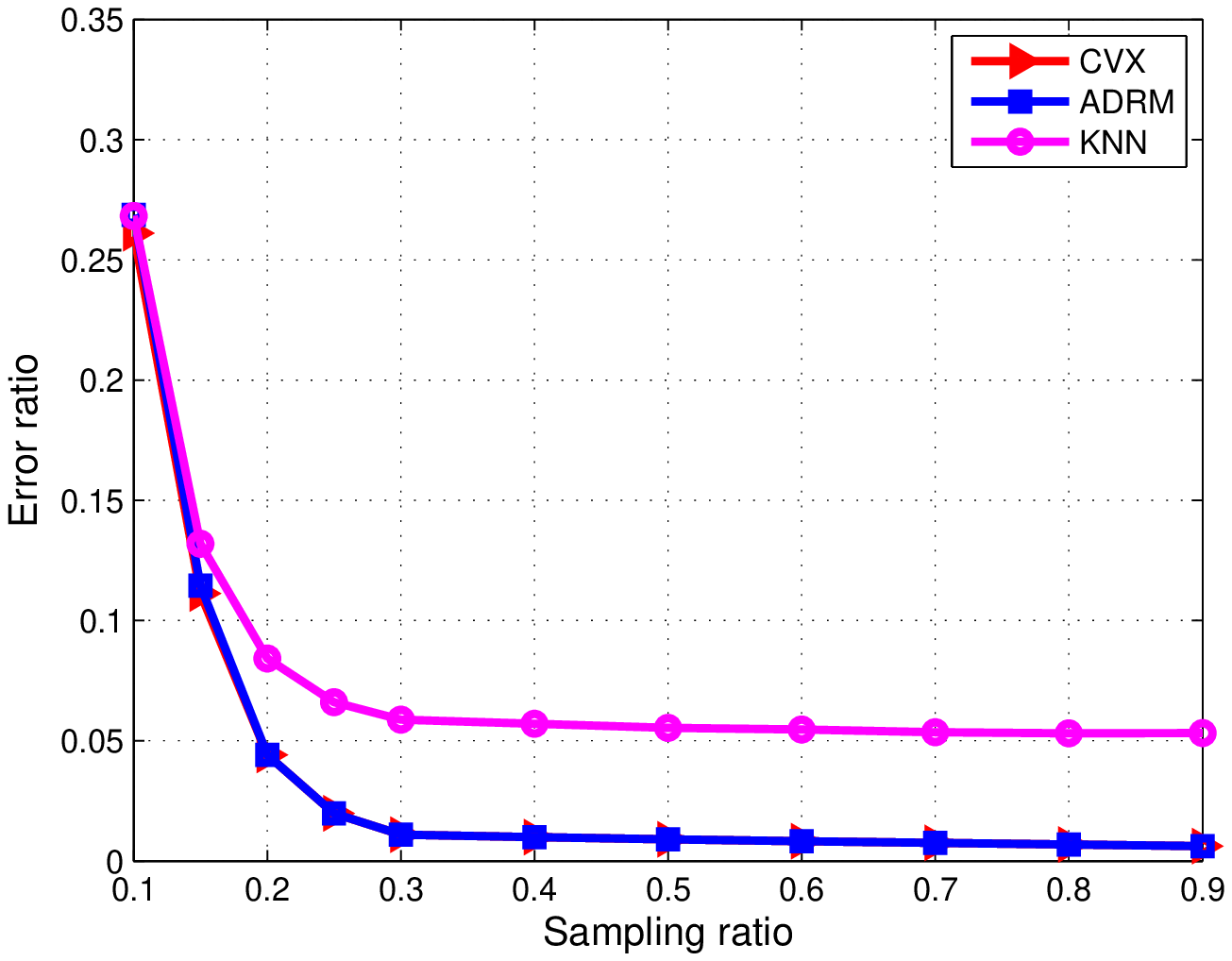}%
\label{fig1_2}}
\hfil
\subfloat[Intel Berkeley (voltage)]{\includegraphics[width=4.4cm]{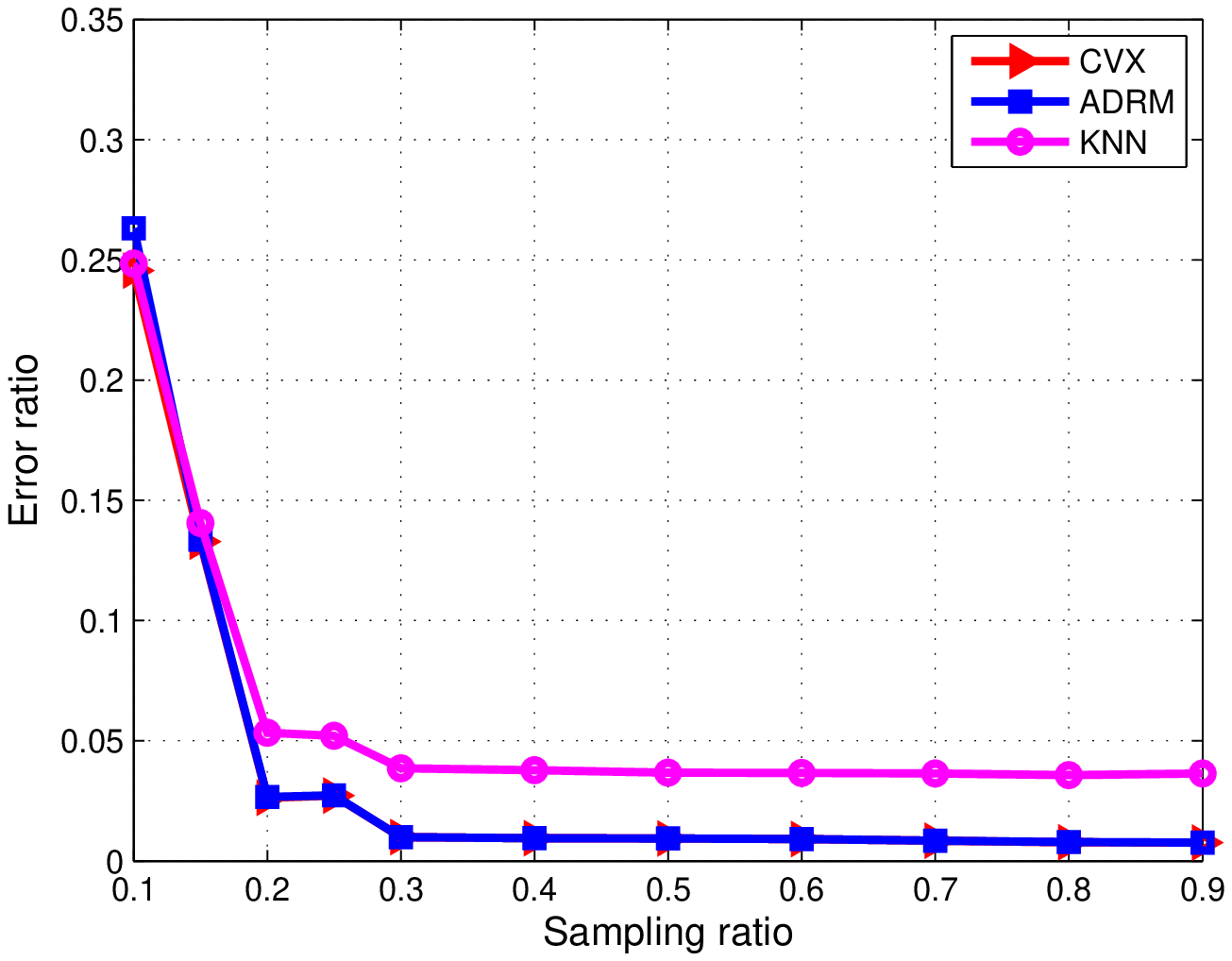}%
\label{fig1_3}}
\hfil
\subfloat[Data Sensing Lab (temperature)]{\includegraphics[width=4.4cm]{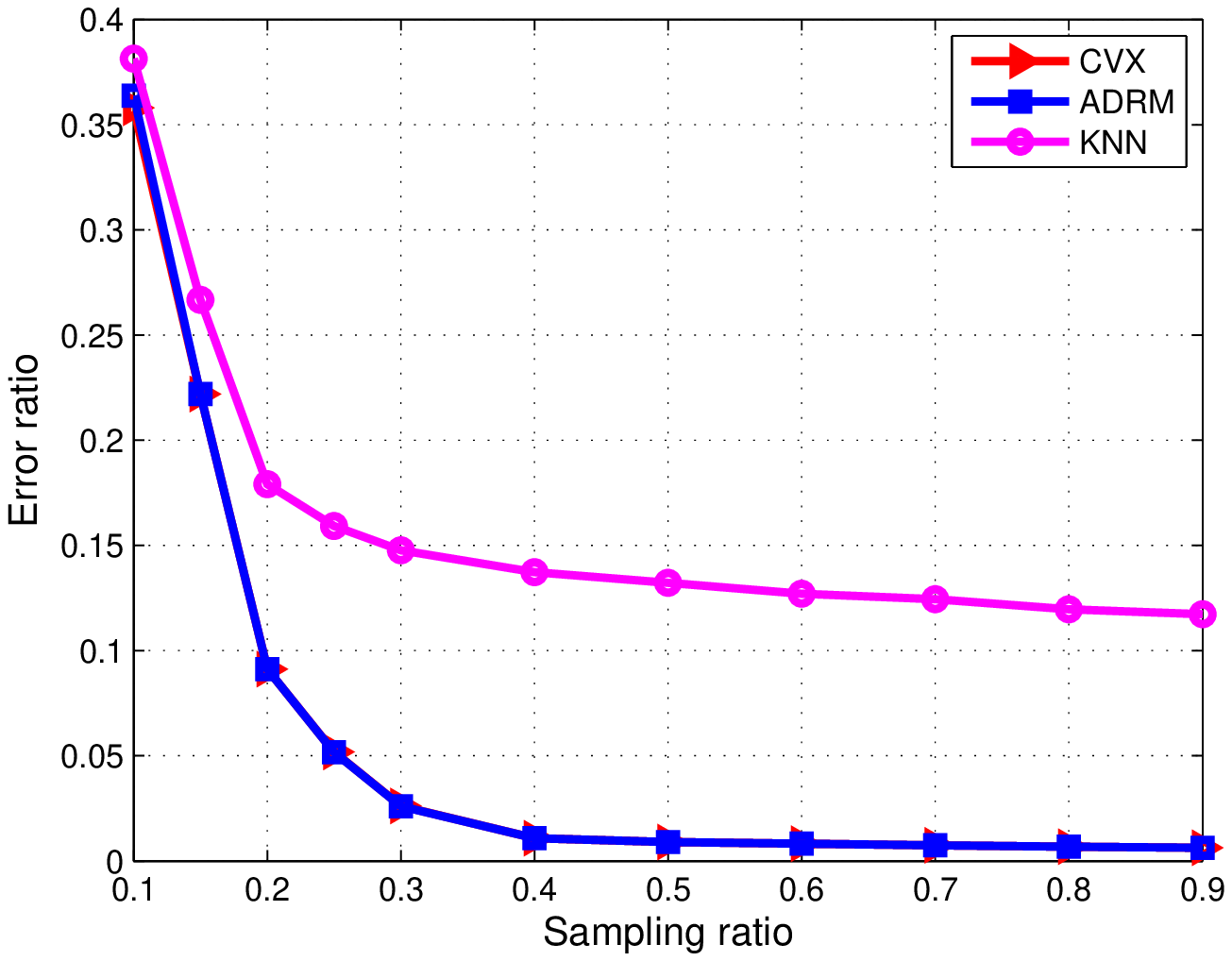}%
\label{fig1_4}}
\hfil
\subfloat[Data Sensing Lab (humidity)]{\includegraphics[width=4.4cm]{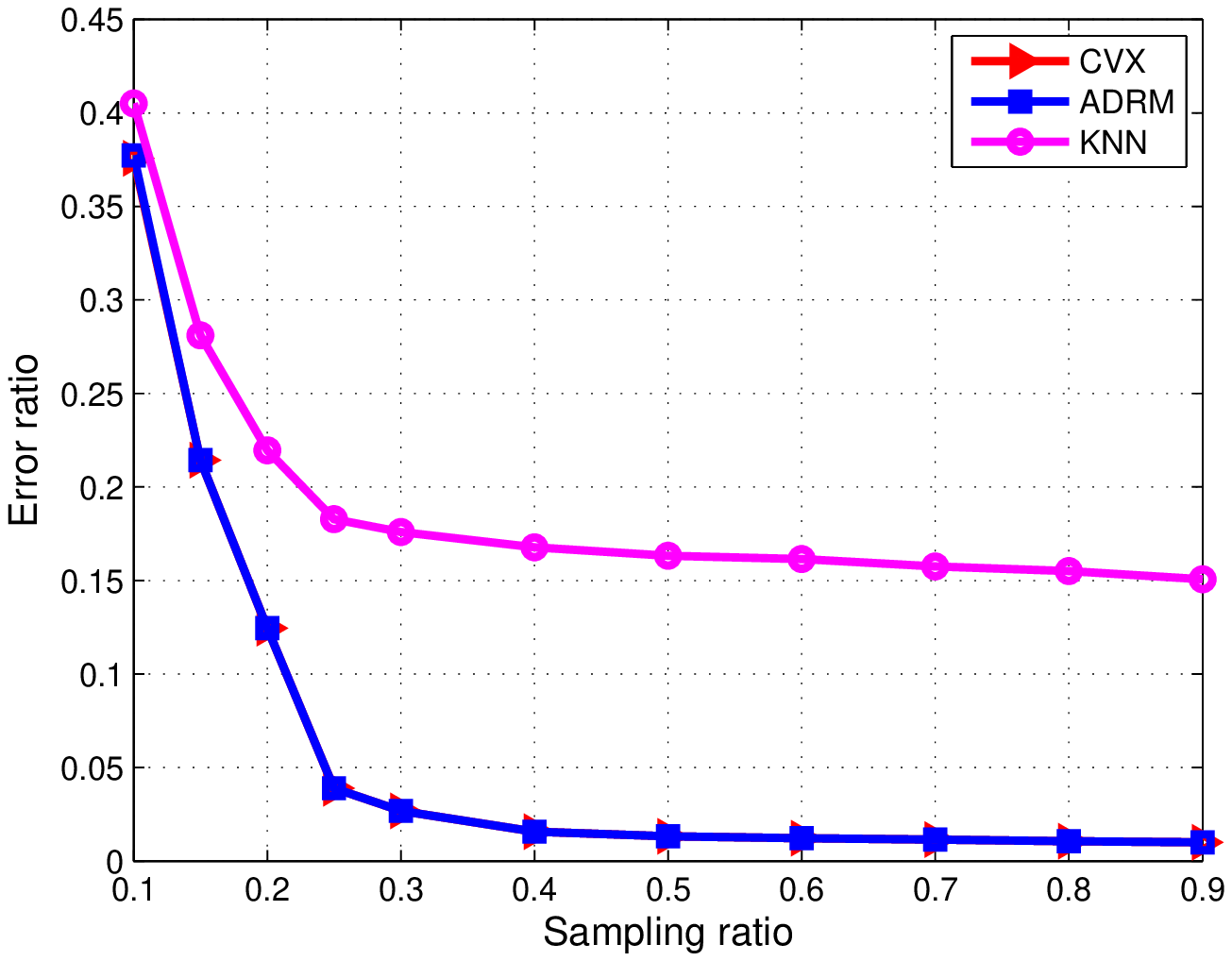}%
\label{fig1_5}}
\hfil
\subfloat[Data Sensing Lab (microphone)]{\includegraphics[width=4.4cm]{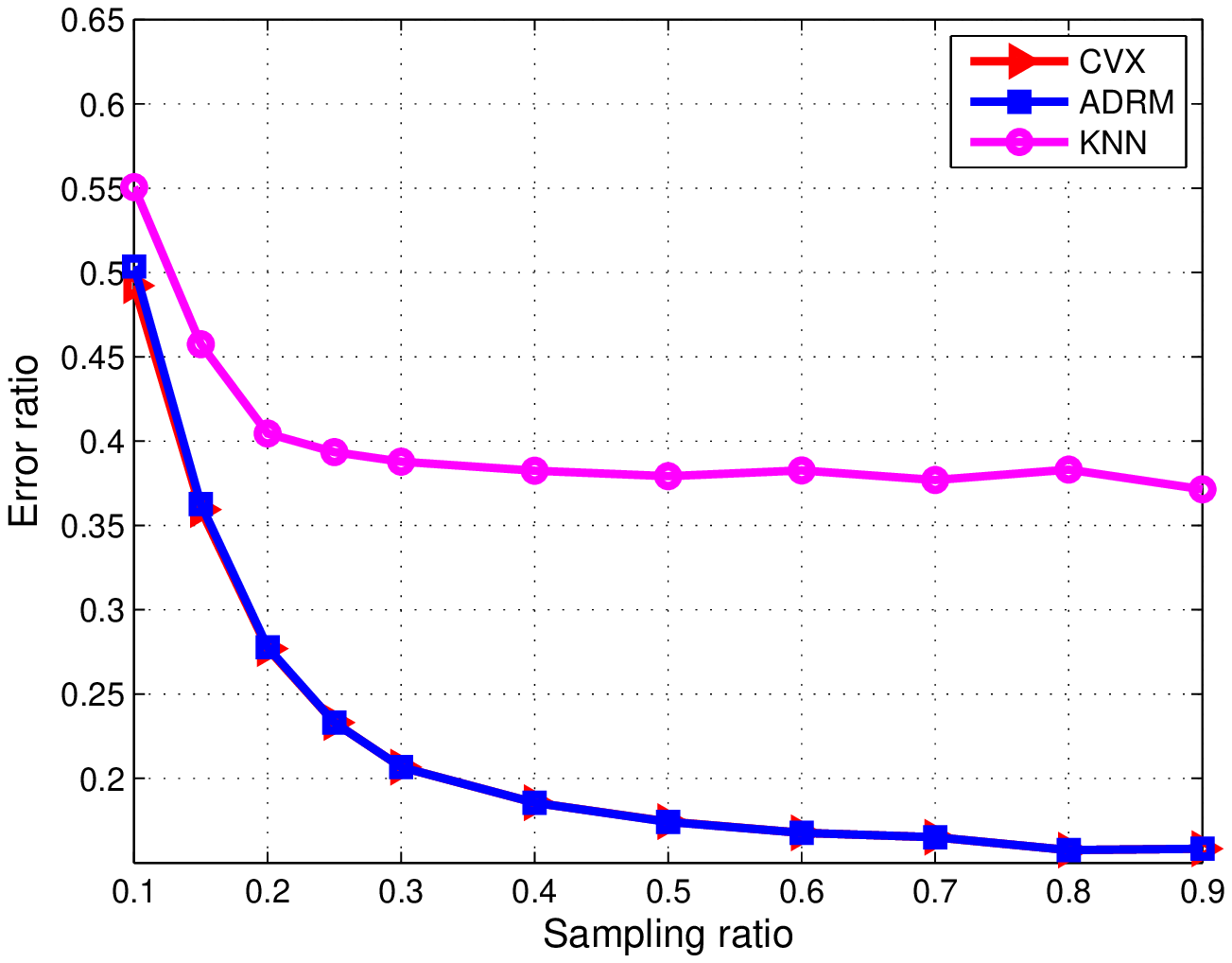}%
\label{fig1_6}}
\caption{Performances of the three algorithms, namely, ADMR, CVX, and KNN, with random missing pattern.}
\label{fig1}
\end{figure}

Fig.~1(a), 1(b), and 1(c) are obtained using the Intel Berkeley dataset. In these figures, ADRM and CVX show the best performance. With roughly $25\%$ of the data, ADRM and CVX can reconstruct all the data with an error ratio of less than $2\%$. In contrast, the error ratio of KNN is close to $5\%$. It can be seen that, with the Intel Berkeley dataste, even KNN can achieve a good performance when sampling ratio is more than $30\%$. This is because that the Intel Berkeley dataset contains indoor data gathered by nodes placed in a 40m$\times$ 30m room, which means that nodes are relatively close to each other. Therefore, the nodes have relatively high spatial correlations.

With the Data Sensing Lab dataset, ADRM and CVX show obvious advantage over KNN compared with the Intel Berkeley dataset, as shown in Fig.~1(d), 1(e), and 1(f). With temperature and humidity data and a sampling ratio of $30\%$, ADRM and CVX can reconstruct the original sensor data with an error ratio of less than $3\%$, whereas that of KNN is closed to $15\%$ and $18\%$, respectively. The reason is that the nodes in data sensing lab are distributed around a conference venue with many separate session rooms and two floors. So the spatial correlation between the nodes is not as strong as that of the nodes in the Intel Berkeley dataset. As a result, KNN performs worse.

It can be seen that all the three algorithms perform poorly with the Data Sensing Lab microphone data. This is because that the sound is much more random, which results in that the microphone data are not in exact low-rank. Yet ADRM and CVX much rely on
the low-rank feature of the original data. Thus, the performance of reconstruction of microphone data is not so good as that of temperature and humidity data. Furthermore, due to the intrinsic property of sound, even neighbouring sensor nodes may gather very different microphone data, which means that the assumption of spatial correlation may fail for microphone data.

In the meantime, from Fig.~1 we can find two interesting points.

Firstly, when sampling ratio exceeds a certain threshold, the error ratios of the three algorithms tend to be steady. For example, with Intel Berkeley temperature data, when sampling ratio exceeds $30\%$, the error ratios of ADRM and CVX remains at about $2\%$, while that of KNN is closed to $5\%$. This indicates that the Intel Berkeley dataset has much redundancy and the whole sensor dataset can be replaced by a small amount of data.

\begin{figure}[!t]
\centering
\subfloat[Intel Berkeley (temperature)]{\includegraphics[width=4.4cm]{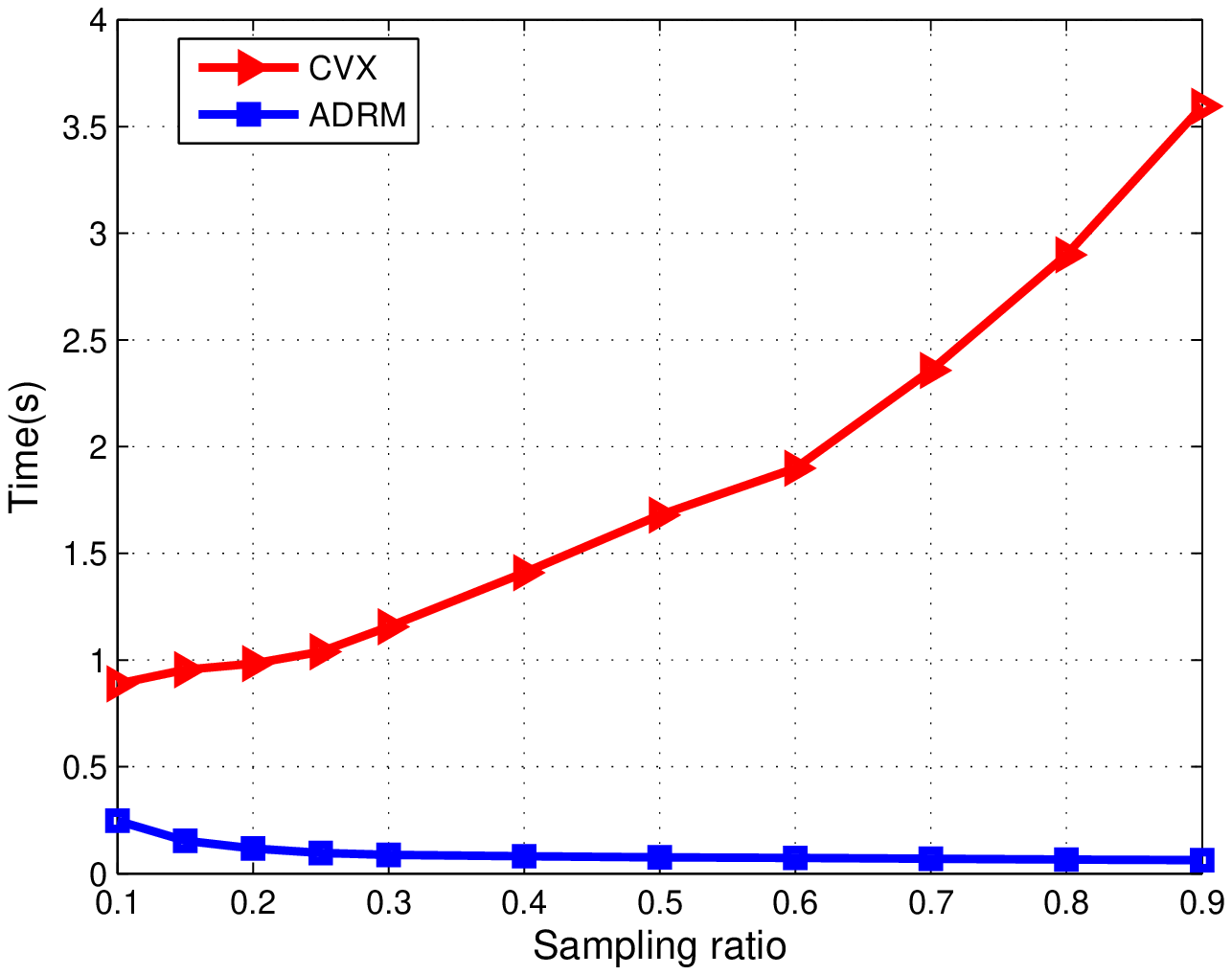}%
\label{fig1}}
\hfil
\subfloat[Data Sensing Lab (temperature)]{\includegraphics[width=4.4cm]{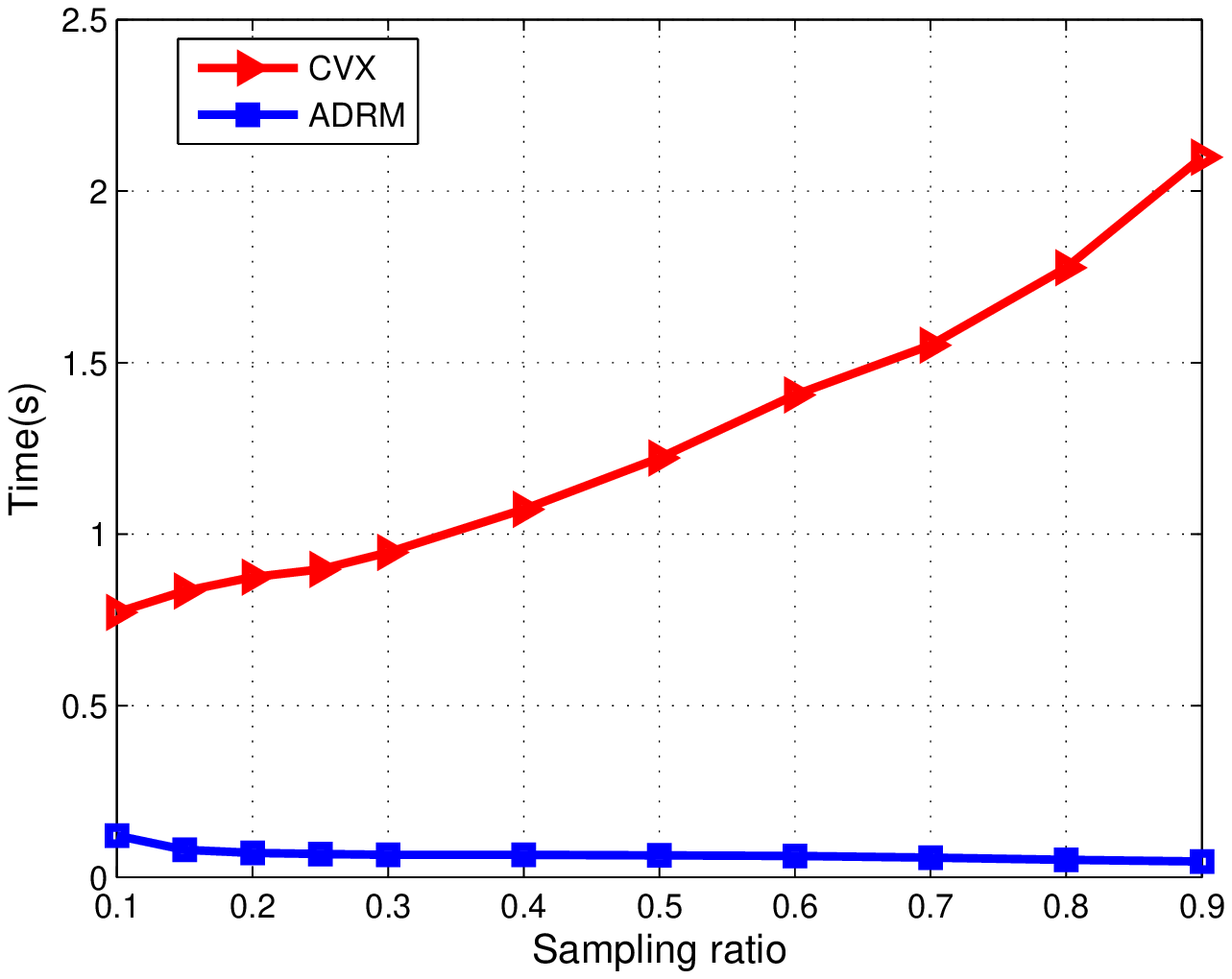}%
\label{fig2}}
\caption{Run time of ADRM and CVX, with random missing pattern.}
\label{fig1}
\end{figure}

Secondly, the error ratios of ADRM and CVX are almost the same. This is because that these two algorithms solve the same convex optimization model (Definition 1) and both achieve the optimal solution. The difference lies in that ADRM is based on the ADMM method which is a first-order method, whereas CVX is based on the interior point method which is a second-order method.
Generally speaking, algorithms based on first-order methods run faster but obtain low- or medium-accuracy solutions. Interior point methods have very high computation complexity and run slower, but they can achieve solutions with higher accuracy. As the error ratio of ADRM is almost the same as that of CVX, we say that our proposed ADRM gets a good reconstruction accuracy. On the other side, from Fig.~2 we can see that ADRM runs evidently faster than CVX. Here we only take the Intel Berkeley temperature data and Data Sensing Lab temperature data as an example. Other sensor data achieve similar results.

In a word, the proposed ADRM algorithm runs faster and obtains optimal solutions with lower error ratios.

\subsubsection{Consecutive Missing Pattern}
\

\begin{figure}[!t]
\centering
\subfloat[Intel Berkeley (temperature)]{\includegraphics[width=4.4cm]{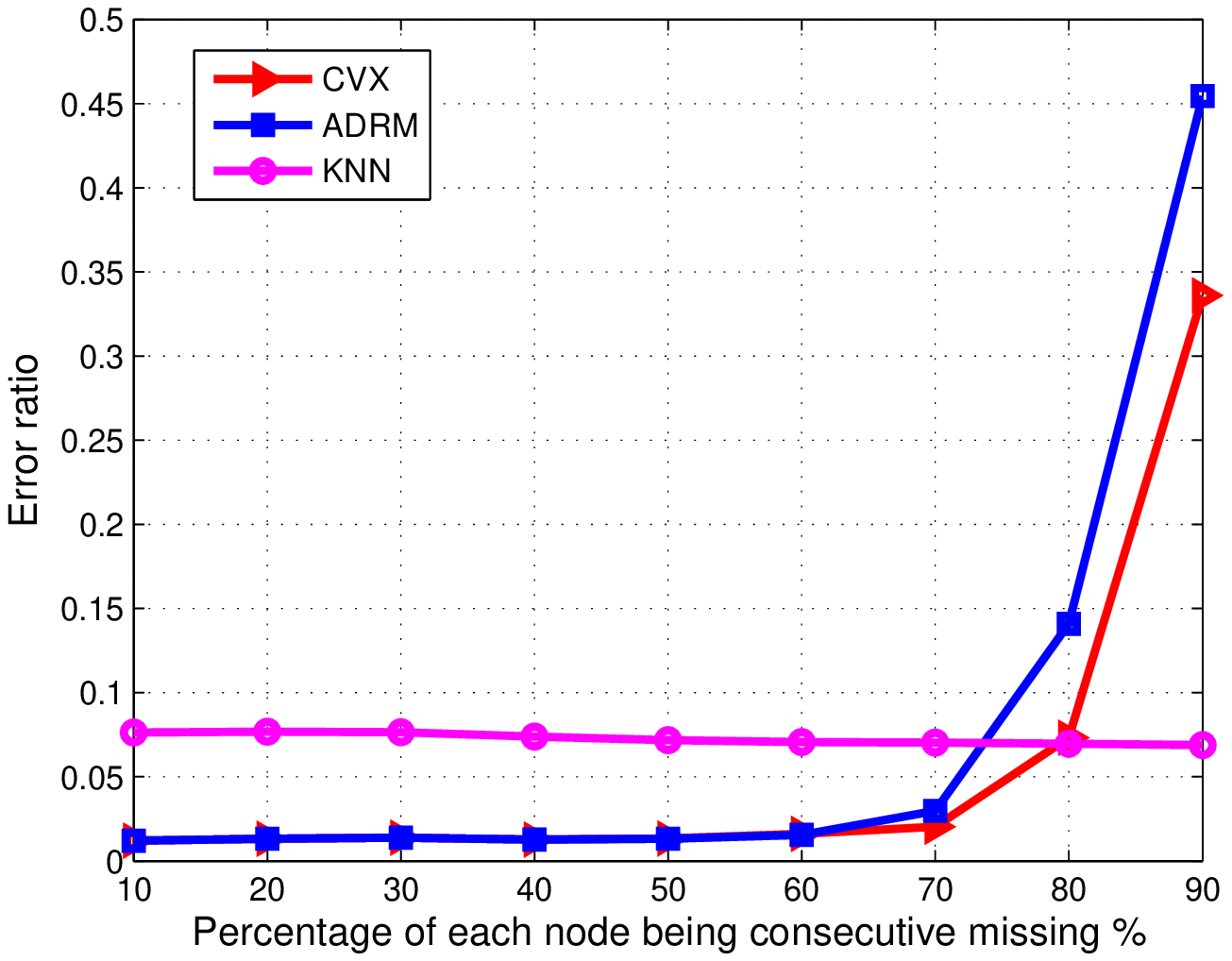}%
\label{fig1}}
\hfil
\subfloat[Intel Berkeley (humidity)]{\includegraphics[width=4.4cm]{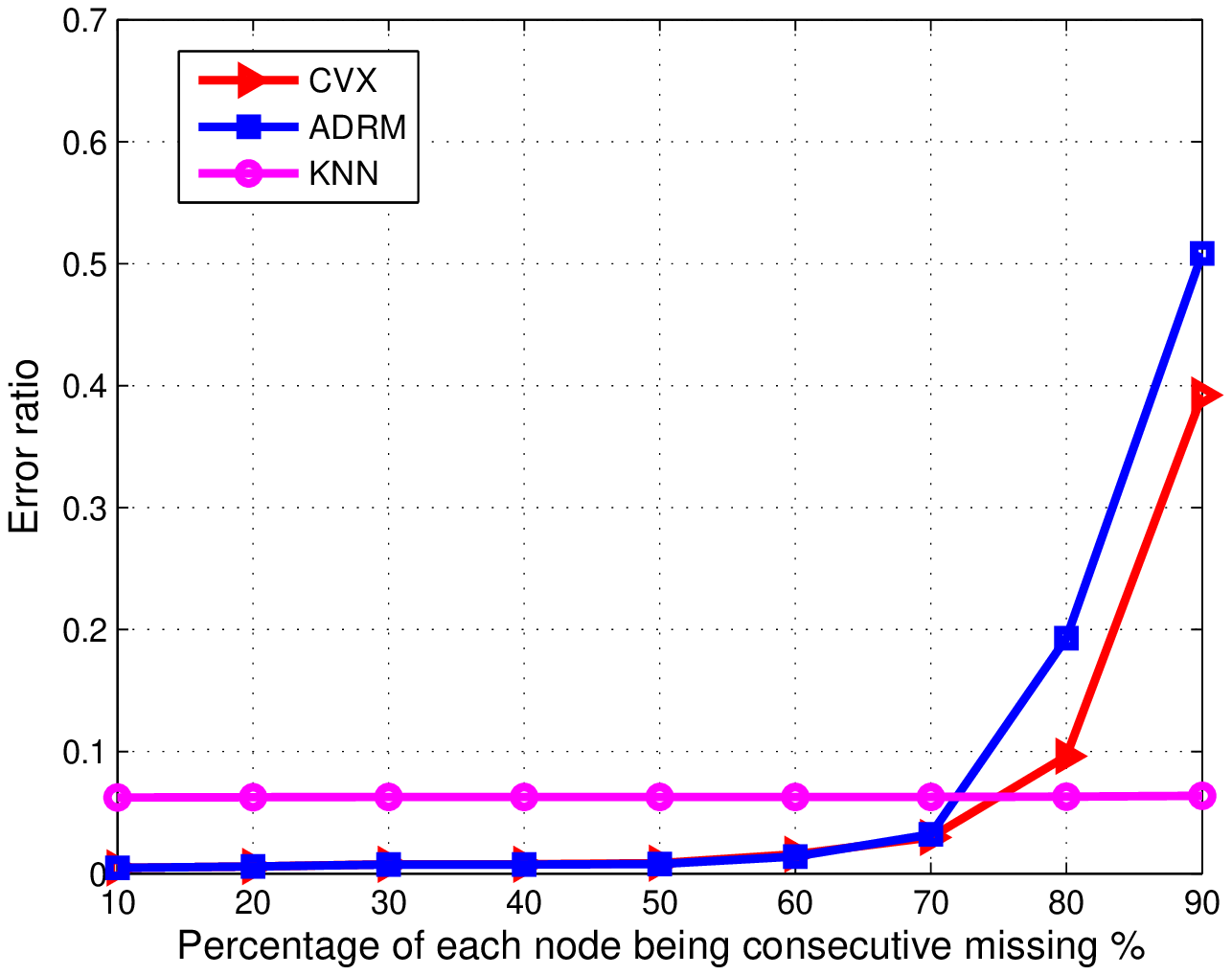}%
\label{fig2}}
\hfil
\subfloat[Intel Berkeley (voltage)]{\includegraphics[width=4.4cm]{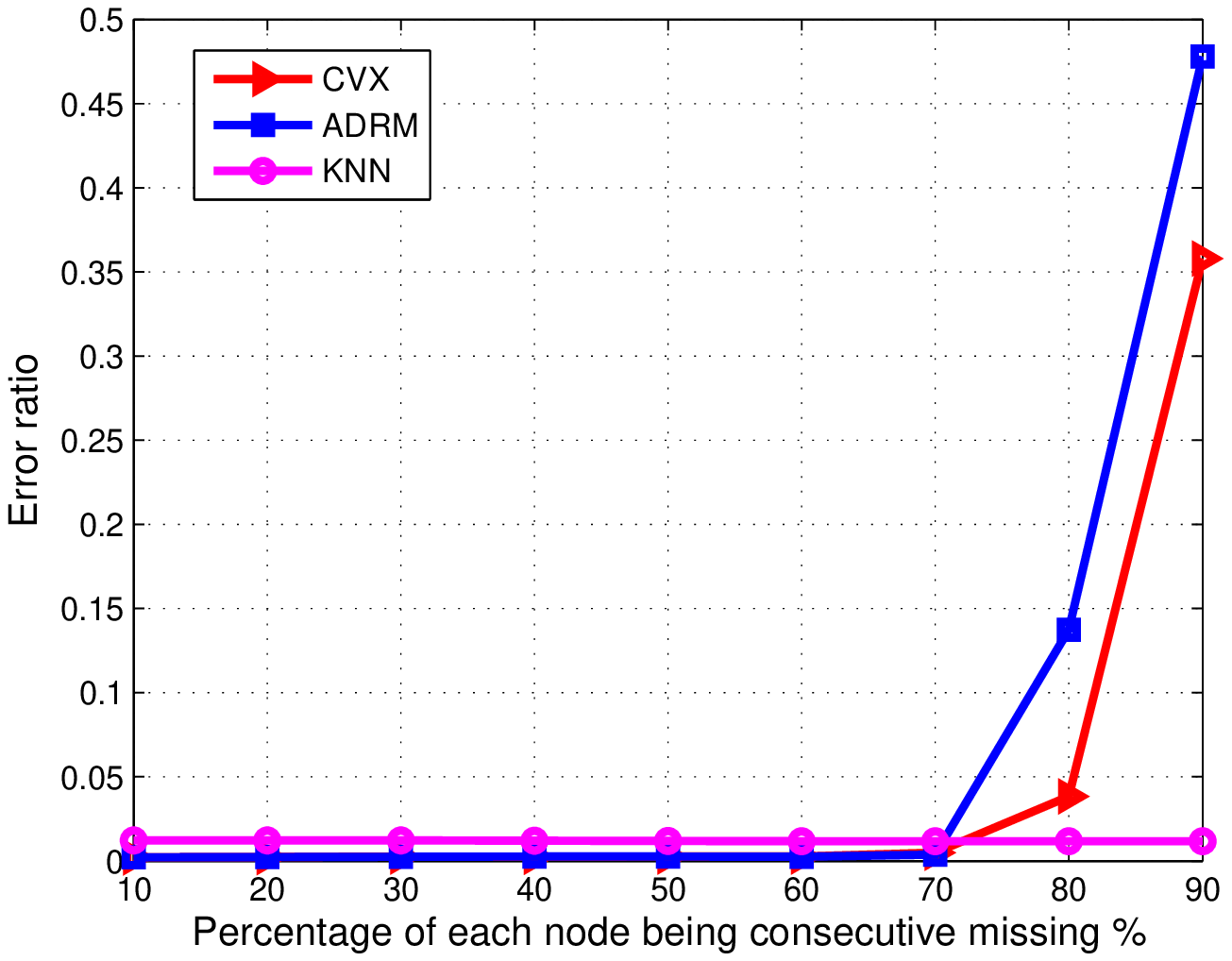}%
\label{fig3}}
\hfil
\subfloat[Data Sensing Lab (temperature)]{\includegraphics[width=4.4cm]{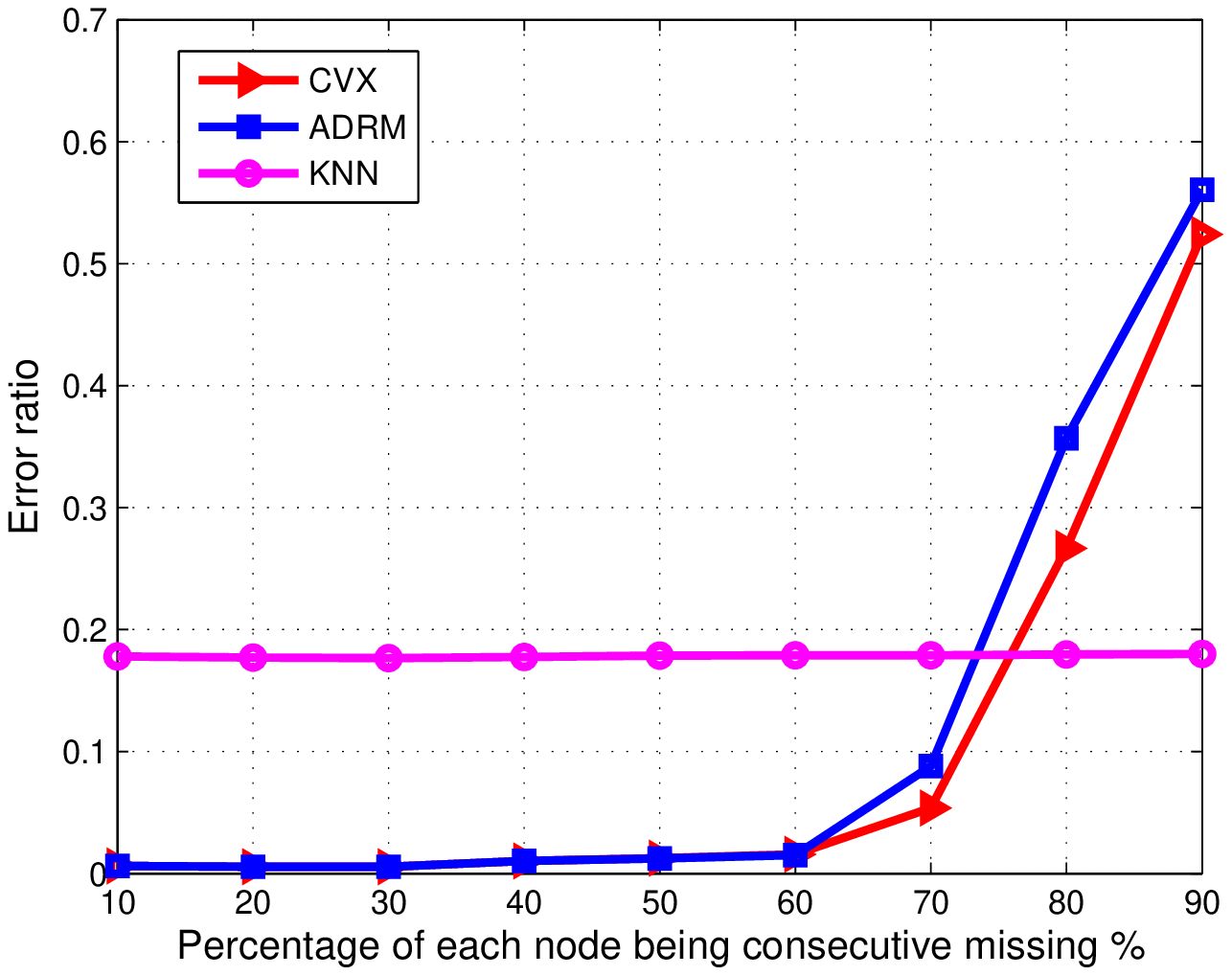}%
\label{fig1.1}}
\hfil
\subfloat[Data Sensing Lab (humidity)]{\includegraphics[width=4.4cm]{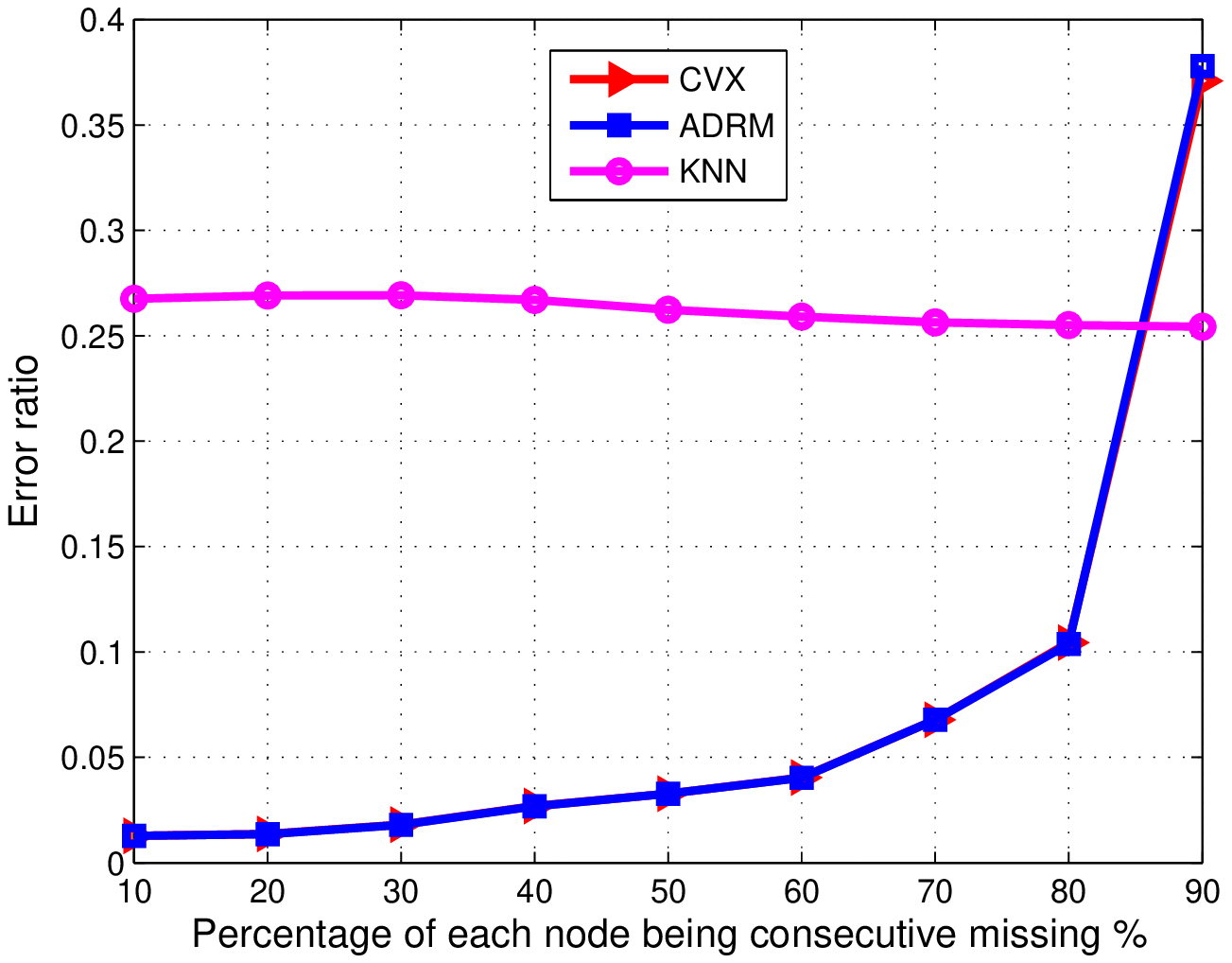}%
\label{fig2.2}}
\hfil
\subfloat[Data Sensing Lab (microphone)]{\includegraphics[width=4.4cm]{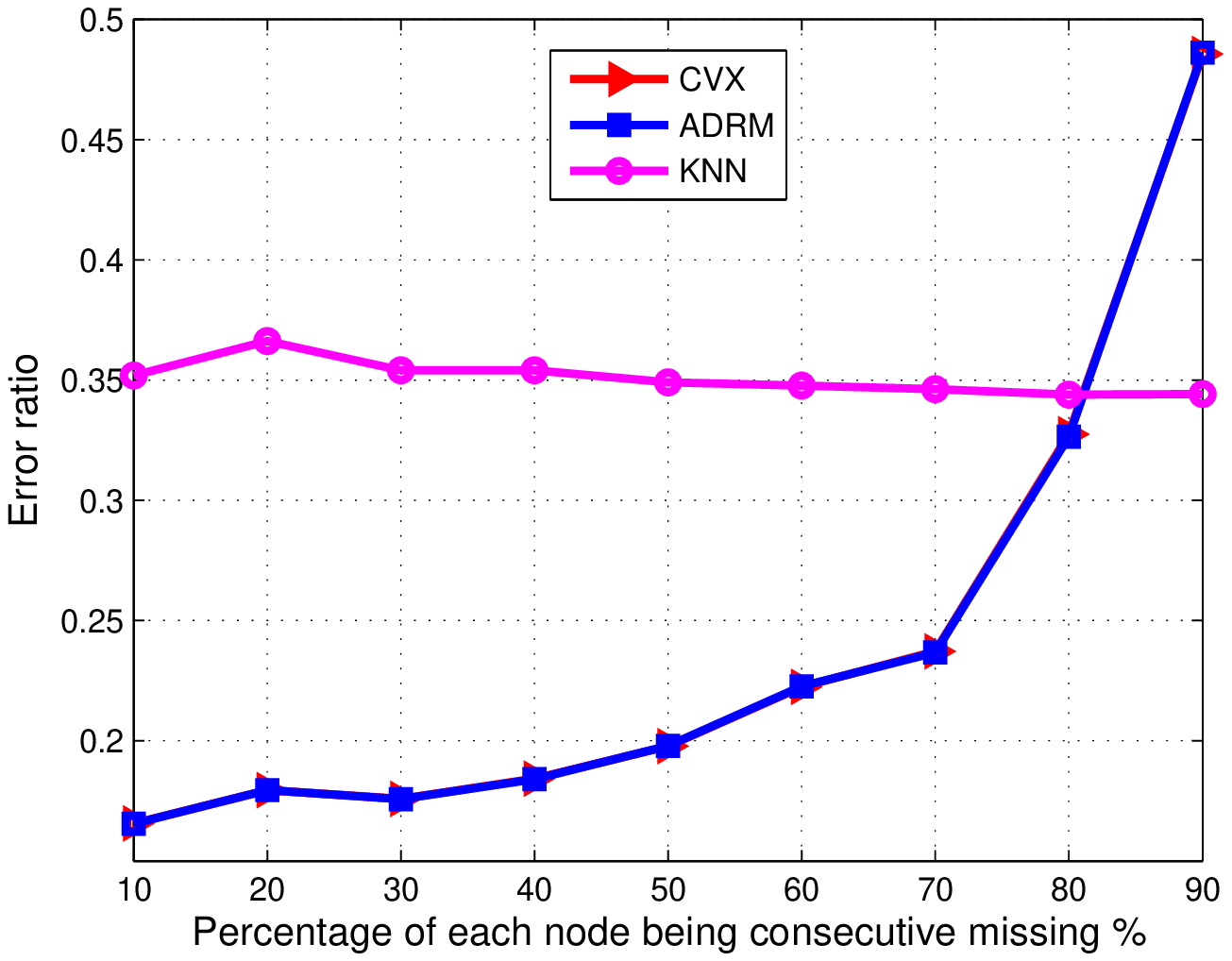}%
\label{fig3.2}}
\caption{Performances of the three algorithms, namely, ADMR, CVX, and KNN, with consecutive missing pattern.}
\label{fig1}
\end{figure}

In this experiment, we calculate error ratios of the three algorithms with consecutive missing pattern, as shown in Fig.~3. The X-axis presents the percentage of consecutive missing.
It can be seen that with consecutive missing pattern,
when the percentage of consecutive missing exceeds $70\%$, the error ratios of ADRM and CVX increase rapidly. However, since KNN utilizes nearby nodes to predict local missing data, the percentage of missing nodes has little impact on error ratio. As a result, the error ratio of KNN remains stable.
For the same reason mentioned in the previous sub-section, all the three algorithms still perform worse with microphone data.

Moreover, we can see that before the turning point, the error ratio of ADRM is almost the same as that of CVX. However, after the turning point, the error ratio of CVX is lower, which means that CVX achieves a higher reconstruction accuracy. The reason is that both ADRM and CVX solve the same convex optimization problem, and before the turning point both of them obtain optimal solutions. But after the turning point, due to the percentage of consecutive missing getting too large, the optimization model of single-attribute sensor-data reconstruction, or Eq.~(8), fails to predict missing data. In this case, the ADRM algorithm, which is based on a first-order method, achieves lower accuracy than CVX.

\begin{figure}[!t]
\centering
\subfloat[Intel Berkeley (humidity)]{\includegraphics[width=4.4cm]{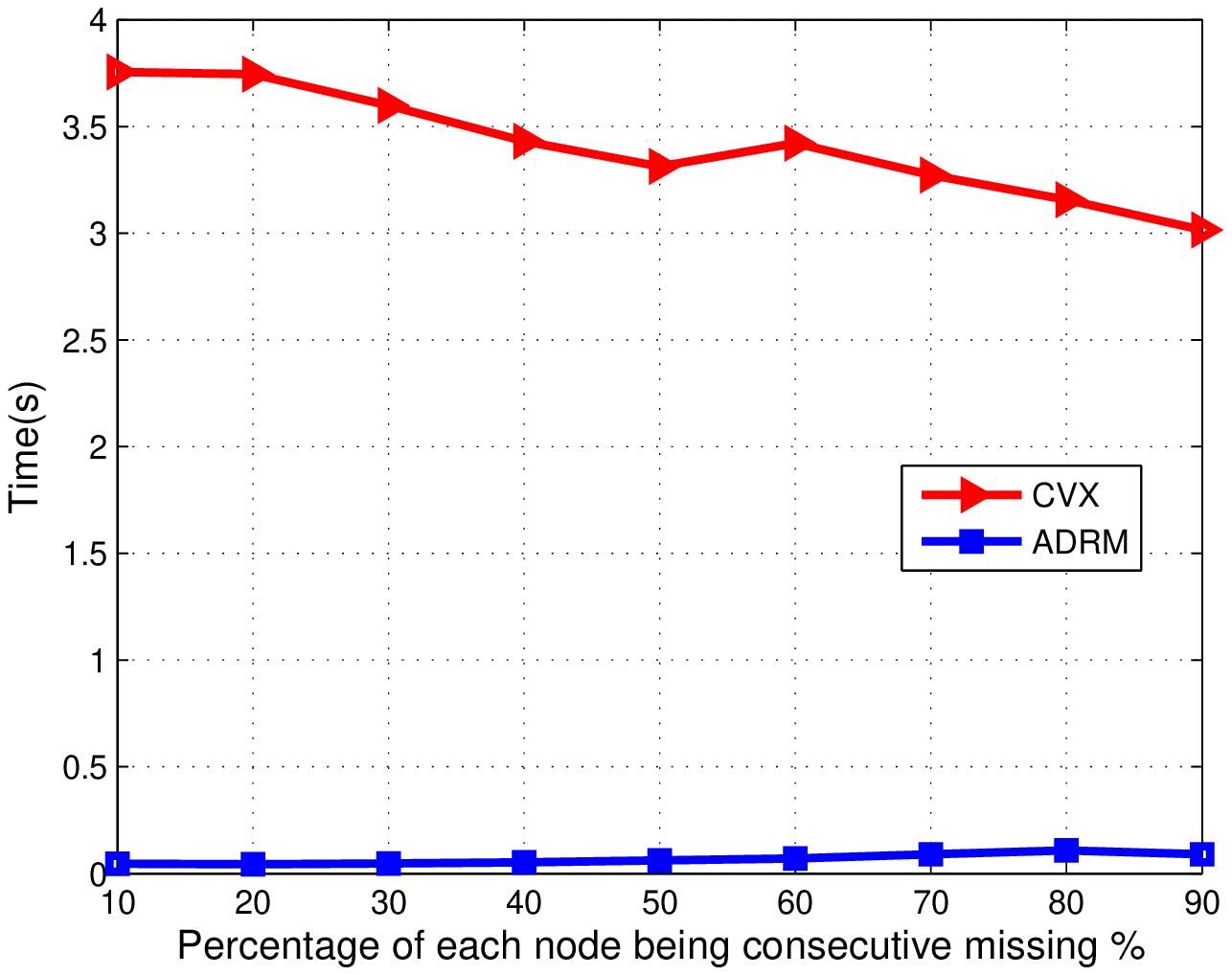}%
\label{fig1}}
\hfil
\subfloat[Data Sensing Lab (humidity)]{\includegraphics[width=4.4cm]{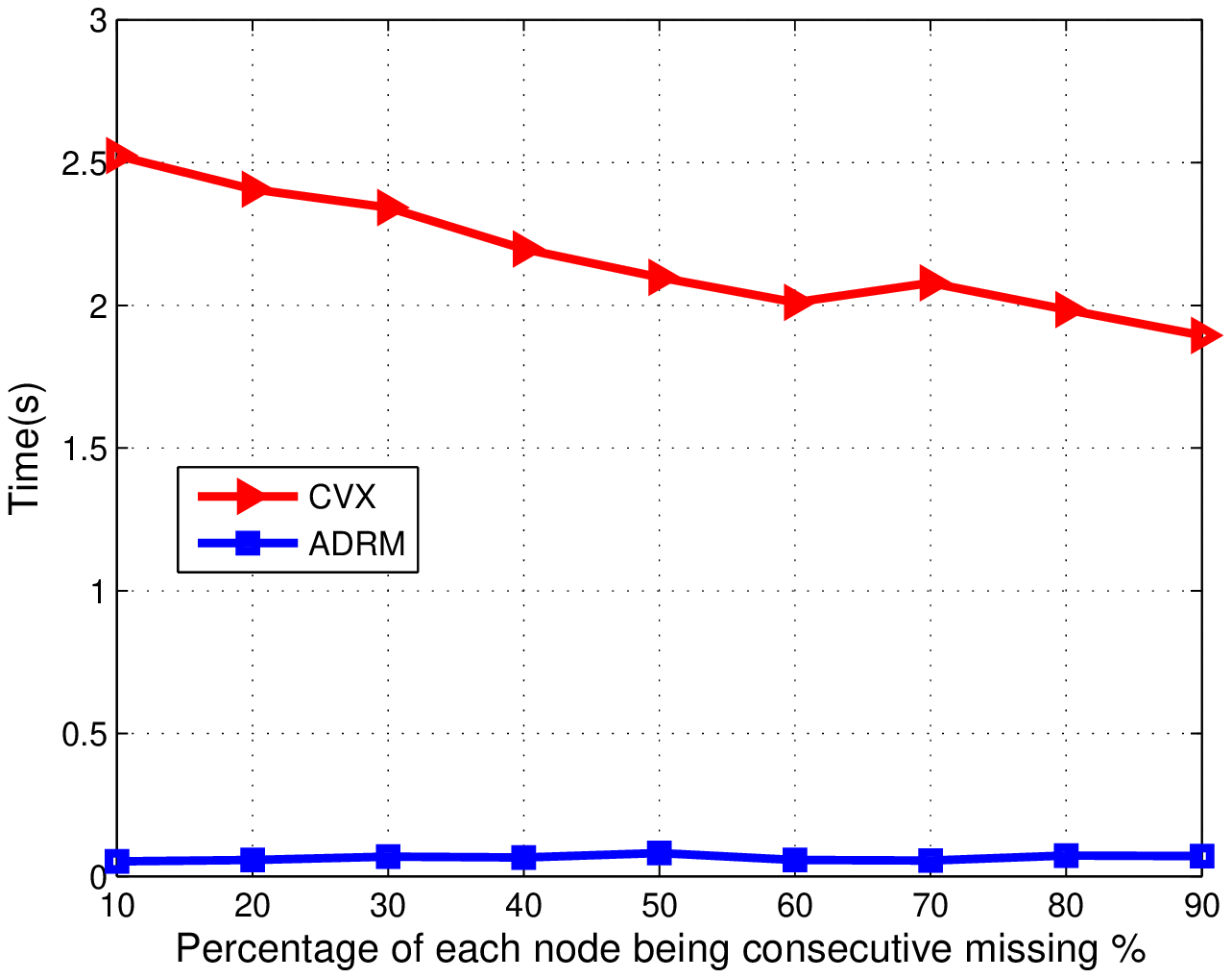}%
\label{fig2}}
\caption{Run time of ADRM and CVX, with consecutive missing pattern.}
\label{fig1}
\end{figure}

From Fig.~4, it can be seen that ADRM runs much faster than CVX. Here we only take Intel Berkeley humidity data and Data Sensing Lab humidity data as an example. Other sensor data achieve similar results.

\subsection{Experiments of Multi-Attribute Sensor-Data Reconstruction}

\begin{figure*}[!t]
\centering
\subfloat[Intel Berkeley dataset]{\includegraphics[width=2.7in]{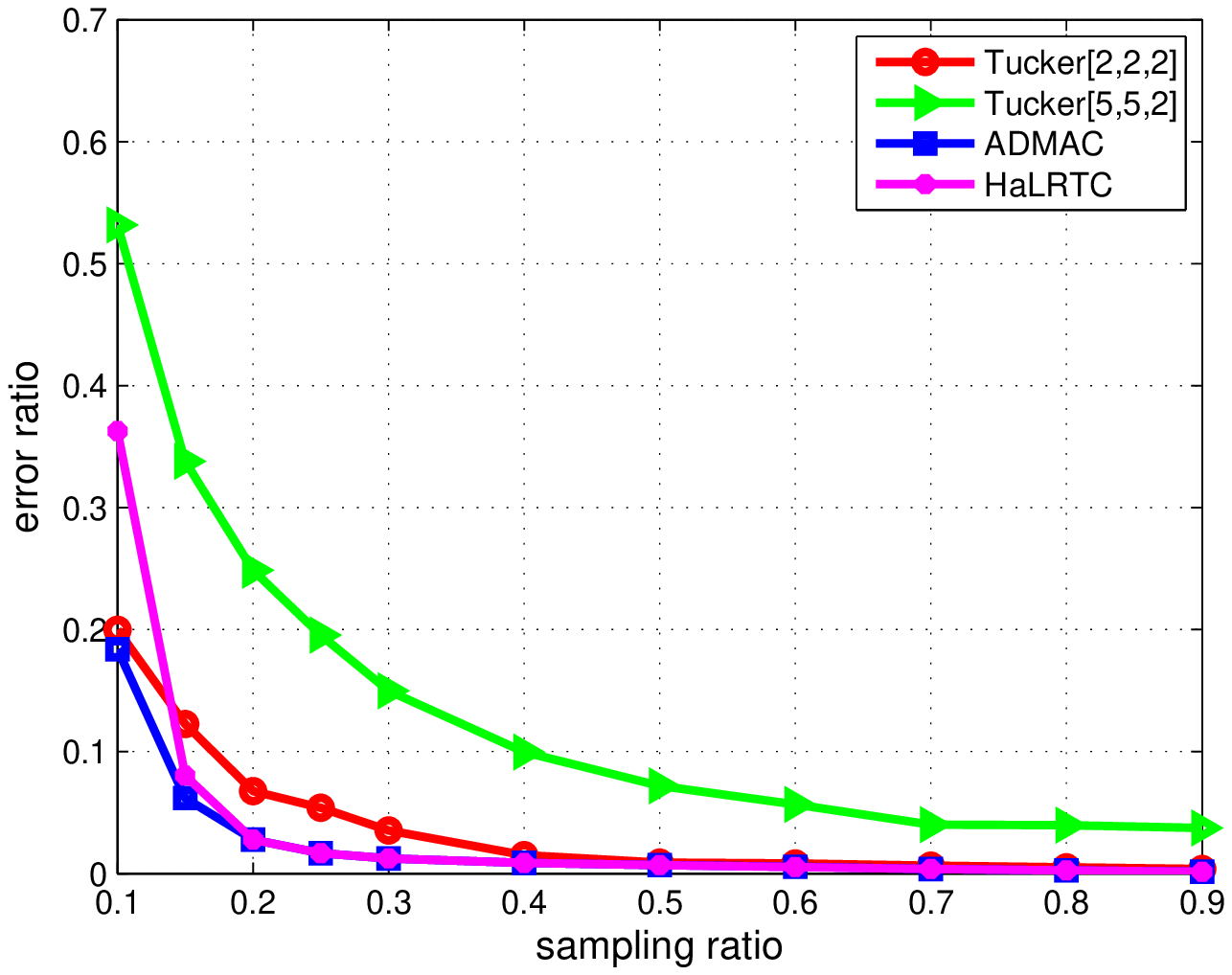}%
\label{fig1}}
\hfil
\subfloat[Data Sensing Lab dataset]{\includegraphics[width=2.7in]{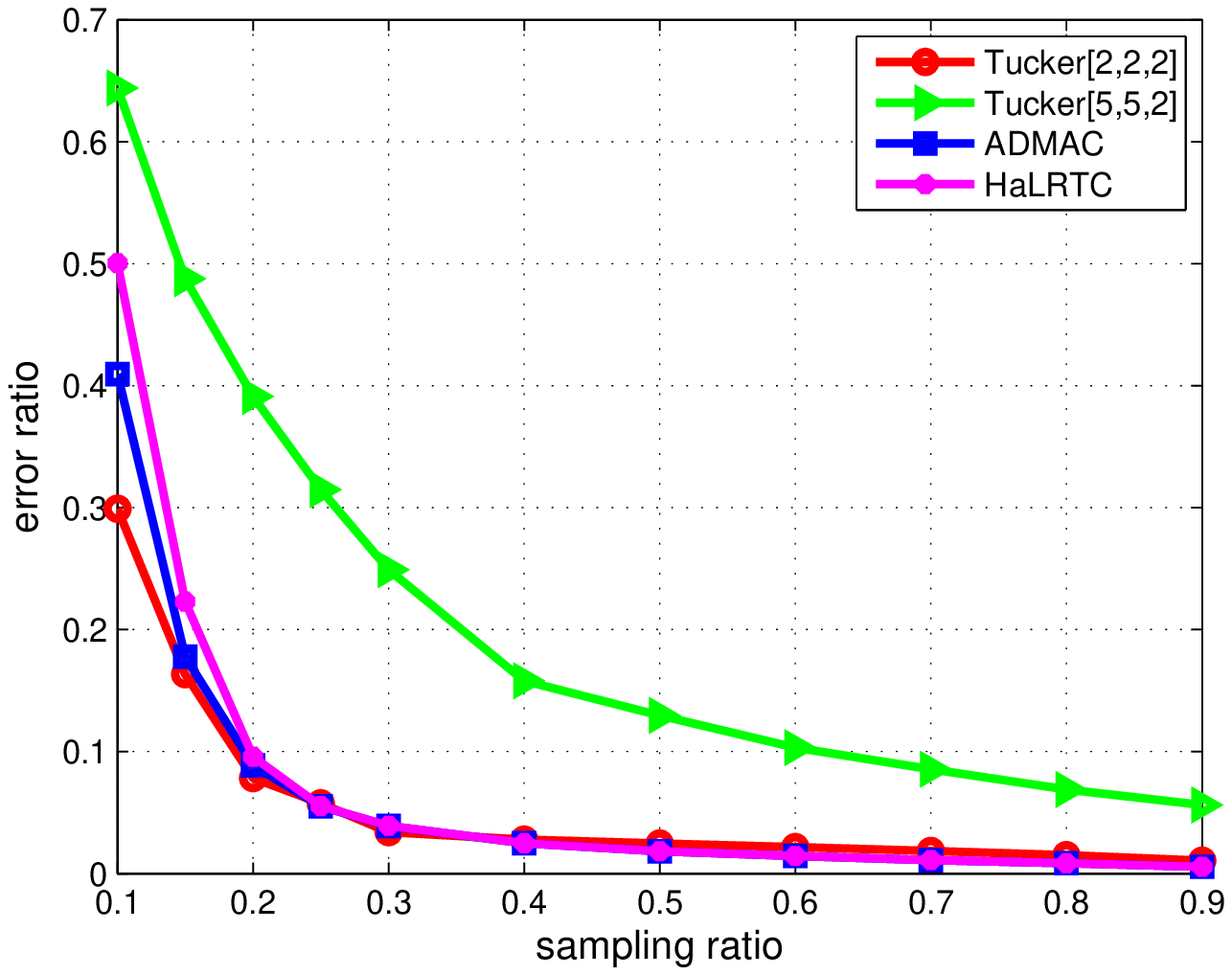}%
\label{fig2}}
\caption{Tensor-based multi-attribute sensor-data reconstruction, with random missing pattern.}
\label{fig5}
\end{figure*}

\begin{figure*}[!t]
\centering
\subfloat[Intel Berkeley dataset]{\includegraphics[width=2.7in]{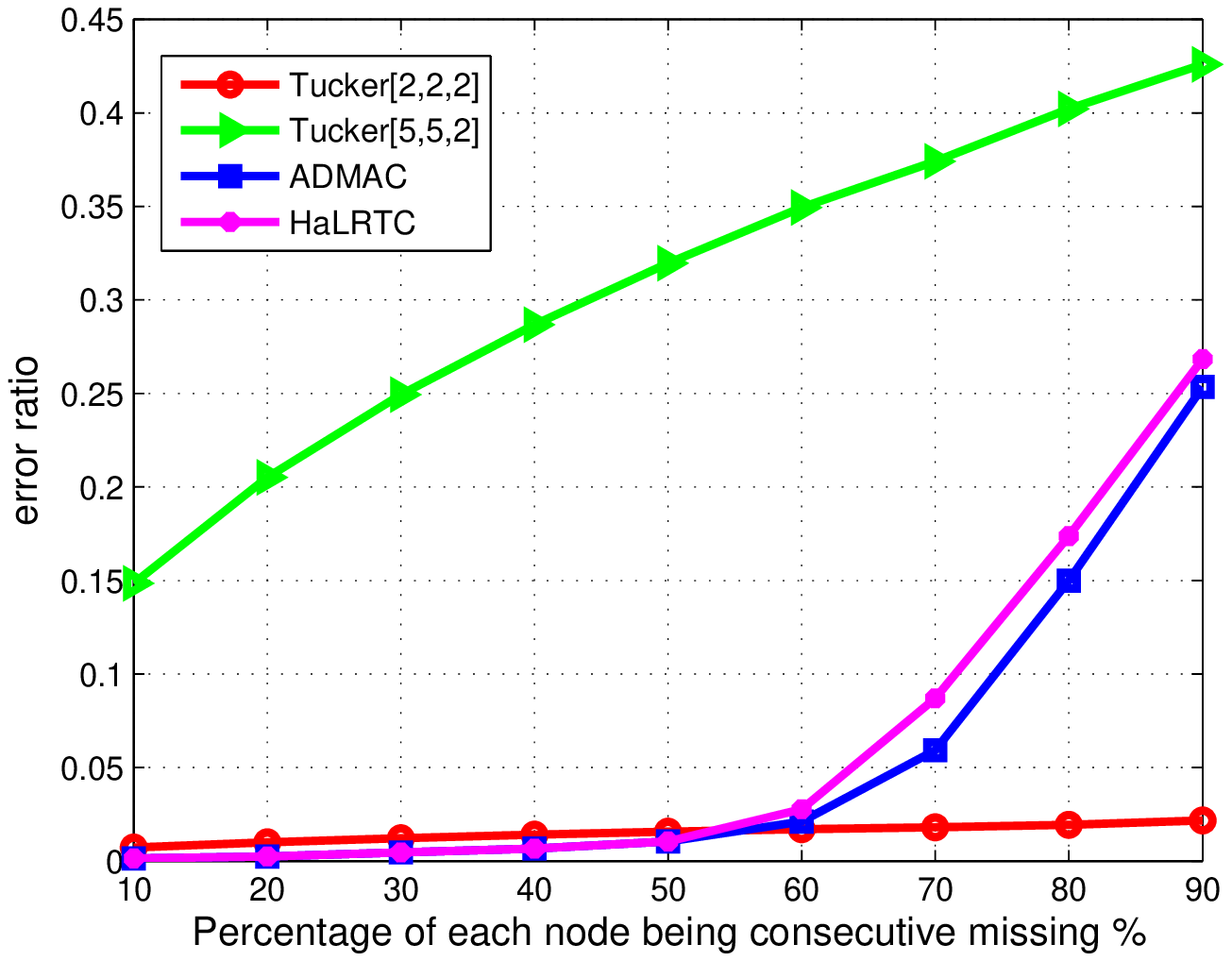}%
\label{fig1}}
\hfil
\subfloat[Data Sensing Lab dataset]{\includegraphics[width=2.7in]{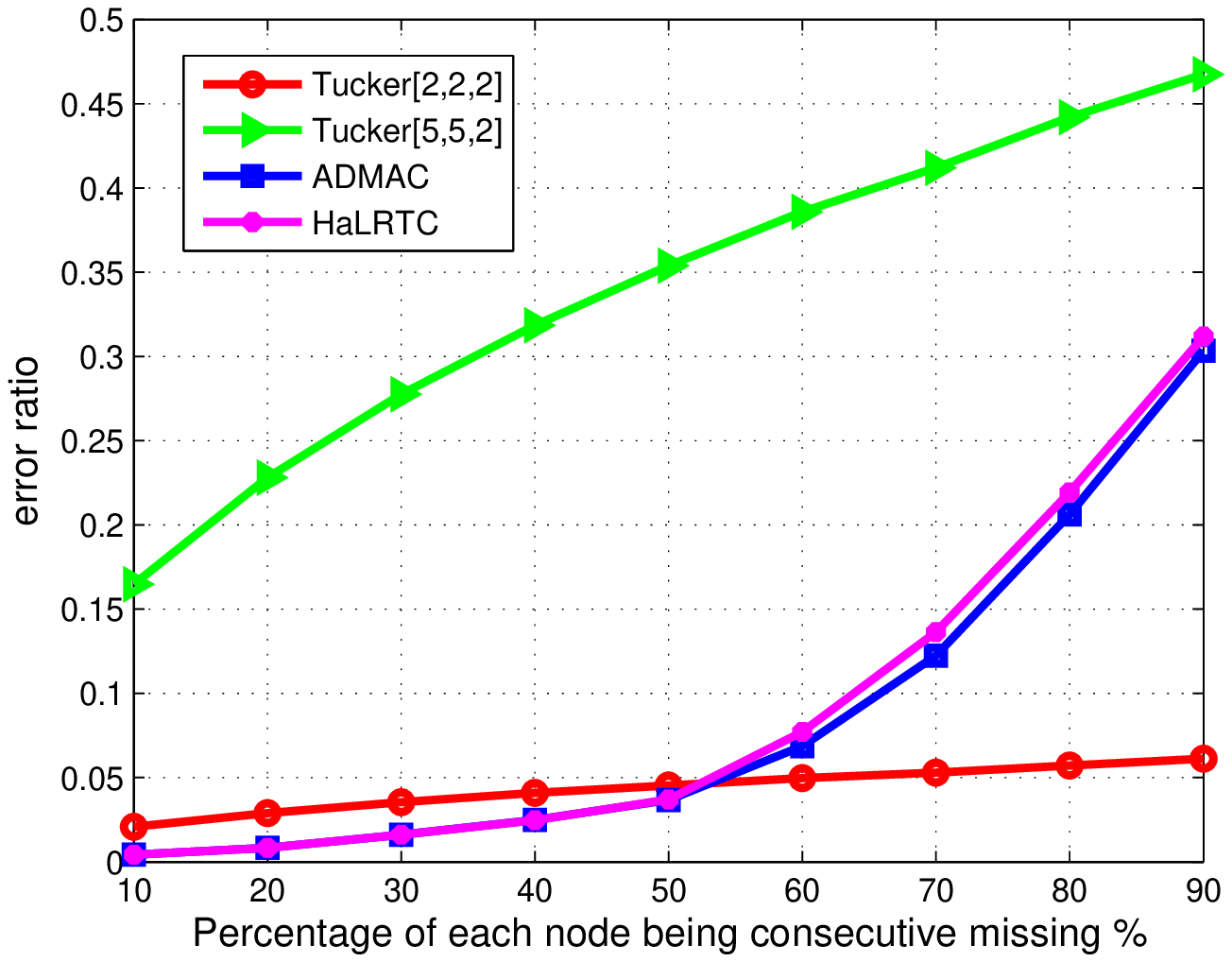}%
\label{fig2}}
\caption{Tensor-based multi-attribute sensor-data reconstruction, with consecutive missing pattern.}
\label{fig6}
\end{figure*}

In this section, we evaluate the performance of the proposed ADMAC algorithm in reconstructing multi-attribute sensor data.
We first use multi-attribute sensor data to constitute a third-order tensor, where the three modes represent sensor time stamp, sensor node ID, and attributes (such as temperature, humidity, and so on), respectively. Thus, we finally obtain a tensor of multi-attribute sensor data.

In this experiment, in order to verify the effectiveness of ADMAC, the ADMAC algorithm is compared with HaLRTC algorithm~\cite{25} and EM-based Tucker decomposition algorithm~\cite{24}. Applying Tucker decomposition to the Intel Berkeley dataset and the Data Sensing Lab dataset, we get Tucker rank of the two constructed tensors. For those datasets, the Tucker rank is approximately rank-[2,2,2]. Then, in contrast, we use the correct rank (rank-[2,2,2]) and a higher rank (here we get rank-[5,5,2]) to do Tucker decomposition.

Fig.~\ref{fig5} shows the results of multi-attribute sensor-data reconstruction using the two datasets with random missing pattern. It can be seen that the proposed ADMAC algorithm performs as good as Tucker decomposition of correct rank-[2,2,2] when sampling ratio is more than $30\%$.
However, with a slightly higher rank-[5,5,2], Tucker decomposition gets poor performance. It means that, in order to use Tucker decomposition based algorithms to accurately reconstruct tensor data, we should first get the correct $n$-rank. However, this is usually intractable in practice, especially when the tensor is incomplete. Moreover, our proposed ADMAC algorithm gets slight advantage over HaLRTC when sampling ratio is less than $20\%$.

Fig.~\ref{fig6} shows the results with consecutive missing pattern. The results are similar to that of random missing pattern. For Tucker decomposition based methods, prior knowledge about $n$-rank of original tensor is critical to reconstruct the tensor with missing data accurately. 

In general, the proposed ADMAC algorithm outperforms other algorithms with the two datasets and the two missing patterns.

\subsection{Experiments of the Relaxed Version of Multi-Attribute Sensor-Data Reconstruction}

\begin{figure}[!t]
\centering
\subfloat[] {\includegraphics[width=2.5in]{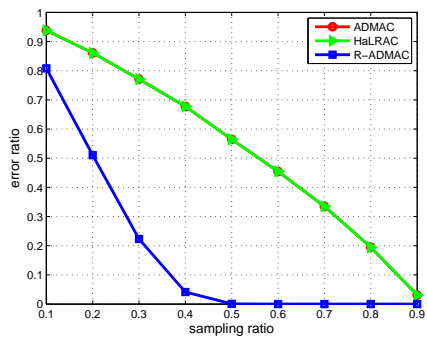}%
\label{fig1}}
\hfil
\subfloat[]{\includegraphics[width=2.5in]{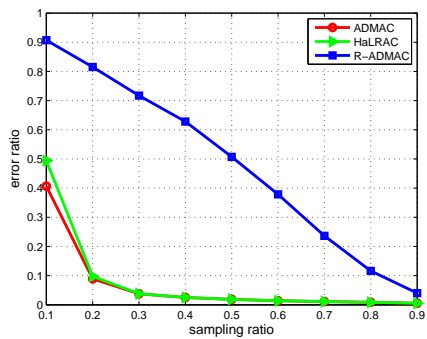}%
\label{fig2}}
\caption{Performance comparison of R-ADMAC, ADMAC, and HaLRTC, with random missing pattern. (a) Random data (a 50$\times$50$\times$50 tensor with a rank of [50,50,5]). (b) Data Sensing Lab dataset.}
\label{fig7}
\end{figure}

In this section, we compare R-ADMAC with ADMAC and HaLRTC. Firstly, we employ simulation to verify the effectiveness of R-ADMAC. We randomly generate a 50$\times50\times50$ tensor with a rank of [50,50,5] and apply R-ADMAC to the generated tensor. From Fig.~7(a) we can see that the R-ADMAC performs the best, whereas ADMAC and HaLRTC perform poorly.

Then we use the Data Sensing Lab dataset to further evaluate R-ADMAC. Fig.~7(b) shows the experiment results. It can be seen that the error ratio of R-ADMAC is large, which means R-ADMAC performs poorly in this case.
It should be noted that R-ADMAC assumes that the original tensor is only low-rank in certain modes, whereas ADMAC and HaLRTC require the original tensor to be jointly low-rank in all modes. That is why R-ADMAC performs poorly with the Data Sensing Lab dataset.
With more and more multiple-attribute sensor data getting available publicly,
we believe the R-ADMAC algorithm will also play an important role for reconstructing multiple-attribute sensor data in practice.

\section{Conclusion}
In this paper, we study the methods of reconstructing missing big sensor data. In order to solve the missing data problem in Internet of things, firstly, we propose a matrix-rank minimization based algorithm, namely, ADRM. ADRM takes full advantage of the low-rank structure of real-world sensor data by computing the minimal low-rank approximations of the incomplete sensor data matrix. Secondly, we consider sensor networks with multiple types of sensors in each node. Accounting for possible correlations among multiple-attribute sensor data, we provide a tensor-based method to estimate missing data and propose an algorithm based on ADMM, namely, ADMAC. ADMAC is based on the assumption that the constructed tensor sensor data is jointly low-rank in all modes. Thirdly, considering that the constructed tensor of multiple-attribute sensor data may not always be low-rank in all modes, we propose a relaxed version of ADMAC, namely, R-ADMAC, which only requires that tensor is low-rank in certain modes.
Finally, we evaluate the algorithms using two real-world sensor network datasets with two missing-data patterns, i.e., random missing pattern and consecutive missing pattern. Experimental results show that the proposed algorithms outperform existing ones.

%

\section*{Acknowledgment}
This work is supported by the Fundamental Research Funds for the Central Universities (N140404015).

\ifCLASSOPTIONcaptionsoff
  \newpage
\fi

\bibliographystyle{IEEEtran}
\bibliography{my}

\begin{thebibliography}{10}
\providecommand{\url}[1]{#1}
\csname url@samestyle\endcsname
\providecommand{\newblock}{\relax}
\providecommand{\bibinfo}[2]{#2}
\providecommand{\BIBentrySTDinterwordspacing}{\spaceskip=0pt\relax}
\providecommand{\BIBentryALTinterwordstretchfactor}{4}
\providecommand{\BIBentryALTinterwordspacing}{\spaceskip=\fontdimen2\font plus
\BIBentryALTinterwordstretchfactor\fontdimen3\font minus
  \fontdimen4\font\relax}
\providecommand{\BIBforeignlanguage}[2]{{%
\expandafter\ifx\csname l@#1\endcsname\relax
\typeout{** WARNING: IEEEtran.bst: No hyphenation pattern has been}%
\typeout{** loaded for the language `#1'. Using the pattern for}%
\typeout{** the default language instead.}%
\else
\language=\csname l@#1\endcsname
\fi
#2}}
\providecommand{\BIBdecl}{\relax}
\BIBdecl

\bibitem{1}
R.~Want, B.~N. Schilit, and S.~Jenson, ``Enabling the internet of things,''
  \emph{Computer}, vol.~48, no.~1, pp. 28--35, 2015.

\bibitem{2}
J.~Gubbi, R.~Buyya, S.~Marusic, and M.~Palaniswami, ``Internet of things
  ({IoT}): A vision, architectural elements, and future directions,''
  \emph{Future Generation Computer Systems}, vol.~29, no.~7, pp. 1645--1660,
  2012.

\bibitem{3}
L.~Mo, Y.~He, Y.~Liu, J.~Zhao, S.~J. Tang, X.~Y. Li, and G.~Dai, ``Canopy
  closure estimates with {GreenOrbs}: Sustainable sensing in the forest,'' in
  \emph{Proceedings of ACM Conference on Embedded Networked Sensor Systems},
  2009, pp. 99--112.

\bibitem{4}
G.~Werner-Allen, K.~Lorincz, J.~Johnson, J.~Lees, and M.~Welsh, ``Fidelity and
  yield in a volcano monitoring sensor network,'' in \emph{Proceedings of
  Symposium on Operating Systems Design and Implementation}, 2006, pp. 27--27.

\bibitem{5}
A.~Ukil, S.~Bandyoapdhyay, C.~Puri, and A.~Pal, ``{IoT} healthcare analytics:
  The importance of anomaly detection,'' in \emph{Proceedings of IEEE
  International Conference on Advanced Information Networking and
  Applications}, 2016, pp. 994--997.

\bibitem{6}
M.~Balazinska, A.~Deshpande, M.~J. Franklin, P.~B. Gibbons, J.~Gray, S.~Nath,
  M.~Hansen, M.~Liebhold, A.~Szalay, and V.~Tao, ``Data management in the
  worldwide sensor web,'' \emph{IEEE Pervasive Computing}, vol.~6, no.~2, pp.
  30--40, 2007.

\bibitem{7}
E.~Granger, M.~A. Rubin, S.~Grossberg, and P.~Lavoie, ``Classification of
  incomplete data using the fuzzy {ARTMAP} neural network,'' in
  \emph{Proceedings of International Joint Conference on Neural Networks},
  2000, pp. 6035--6035.

\bibitem{8}
T.~Cover and P.~Hart, ``Nearest neighbor pattern classification,'' \emph{IEEE
  Transactions on Information Theory}, vol.~13, no.~1, pp. 21--27, 1967.

\bibitem{9}
M.~Halatchev and G.~Le, ``Estimating missing values in related sensor data
  streams,'' in \emph{Proceedings of International Conference on Management of
  Data}, 2005, pp. 83--94.

\bibitem{10}
L.~Gruenwald, H.~Chok, and M.~Aboukhamis, ``Using data mining to estimate
  missing sensor data,'' in \emph{Proceedings of IEEE International Conference
  on Data Mining Workshops}, 2007, pp. 207--212.

\bibitem{11}
C.~Alippi, G.~Boracchi, and M.~Roveri, ``On-line reconstruction of missing data
  in sensor/actuator networks by exploiting temporal and spatial redundancy,''
  in \emph{Proceedings of International Joint Conference on Neural Networks},
  2012, pp. 1--8.

\bibitem{12}
Y.~Li, C.~Ai, W.~P. Deshmukh, and Y.~Wu, ``Data estimation in sensor networks
  using physical and statistical methodologies,'' in \emph{Proceedings of
  International Conference on Distributed Computing Systems}, 2008, pp.
  538--545.

\bibitem{13}
L.~Pan and J.~Li, ``K-nearest neighbor based missing data estimation algorithm
  in wireless sensor networks,'' \emph{Wireless Sensor Network}, vol.~2, no.~2,
  pp. 115--122, 2010.

\bibitem{14}
C.~Y. Li, W.~L. Su, T.~G. Mckenzie, and F.~C. Hsu, ``Recommending missing
  sensor values,'' in \emph{Proceedings of IEEE International Conference on Big
  Data}, 2015, pp. 381--390.

\bibitem{15}
X.~Wu, C.~L. Chuang, and J.~A. Jiang, ``Temperature map recovery based on
  compressive sensing for large-scale wireless sensor networks,'' in
  \emph{Proceedings of Green Computing and Communications}, 2013, pp. 1202 --
  1206.

\bibitem{16}
J.~Cheng, H.~Jiang, X.~Ma, L.~Liu, L.~Qian, C.~Tian, and W.~Liu, ``Efficient
  data collection with sampling in {WSNs}: Making use of matrix completion
  techniques,'' in \emph{Proceedings of IEEE GLOBECOM}, 2010, pp. 1--5.

\bibitem{17}
X.~Piao, Y.~Hu, Y.~Sun, and B.~Yin, ``Efficient data gathering in wireless
  sensor networks based on low rank approximation,'' in \emph{Proceedings of
  Green Computing and Communications}, 2013, pp. 699--706.

\bibitem{18}
J.~Cheng, Q.~Ye, H.~Jiang, D.~Wang, and C.~Wang, ``{STCDG}: An efficient data
  gathering algorithm based on matrix completion for wireless sensor
  networks,'' \emph{IEEE Transactions on Wireless Communications}, vol.~12,
  no.~2, pp. 850--861, 2013.

\bibitem{19}
L.~Kong, M.~Xia, X.~Y. Liu, M.~Y. Wu, and X.~Liu, ``Data loss and
  reconstruction in sensor networks,'' in \emph{Proceedings of IEEE INFOCOM},
  2013, pp. 1654--1662.

\bibitem{20}
G.~Chen, X.~Y. Liu, L.~Kong, J.~L. Lu, Y.~Gu, W.~Shu, and M.~Y. Wu, ``Multiple
  attributes-based data recovery in wireless sensor networks,'' in
  \emph{Proceeding of IEEE GLOBECOM}, 2013, pp. 103--108.

\bibitem{21}
L.~Kong, M.~Xia, X.~Y. Liu, G.~Chen, Y.~Gu, M.~Y. Wu, and X.~Liu, ``Data loss
  and reconstruction in wireless sensor networks,'' \emph{IEEE Transaction on
  Parallel and Distributed Systems}, vol.~25, no.~11, pp. 2818--2828, 2014.

\bibitem{22}
G.~Chen, X.~Y. Liu, L.~Kong, and J.~L. Lu, ``{JSSDR}: Joint-sparse sensory data
  recovery in wireless sensor networks,'' in \emph{Proceedings of Wireless and
  Mobile Computing, Networking and Communications}, 2013, pp. 367--374.

\bibitem{23}
S.~Gandy, B.~Recht, and I.~Yamada, ``Tensor completion and low-n-rank tensor
  recovery via convex optimization,'' \emph{Inverse Problems}, vol.~27, no.~2,
  pp. 25\,010--25\,028(19), 2011.

\bibitem{25}
J.~Liu, P.~Musialski, P.~Wonka, and J.~Ye, ``Tensor completion for estimating
  missing values in visual data,'' \emph{IEEE Transactions on Pattern Analysis
  and Machine Intelligence}, vol.~35, no.~1, pp. 208--220, 2013.

\bibitem{26}
E.~J. Cand$\grave{e}$s and B.~Recht, ``Exact matrix completion via convex
  optimization,'' \emph{Foundations of Computational Mathematics}, vol.~9,
  no.~6, pp. 717--772, 2008.

\bibitem{27}
M.~Grant and S.~Boyd, ``{CVX}: Matlab software for disciplined convex
  programming, version 2.1,'' \url{http://cvxr.com/cvx}, Mar. 2014.

\bibitem{30}
S.~Boyd, N.~Parikh, E.~Chu, B.~Peleato, and J.~Eckstein, ``Distributed
  optimization and statistical learning via the alternating direction method of
  multipliers,'' \emph{Foundations and Trends in Machine Learning}, vol.~3,
  no.~1, pp. 1--122, 2011.

\bibitem{31}
J.~F. Cai, C.~E. J., and Z.~Shen, ``A singular value thresholding algorithm for
  matrix completion,'' \emph{Siam Journal on Optimization}, vol.~20, no.~4, pp.
  1956--1982, 2010.

\bibitem{32}
{Data Sensing Lab dataset}, Available:\url{http://datasensinglab.com/}.

\bibitem{34}
T.~G. Kolda and B.~W. Bader, ``Tensor decompositions and applications,''
  \emph{College and Research Libraries}, vol.~66, no.~4, pp. 294--310, 2005.

\bibitem{35}
R.~Tomioka, K.~Hayashi, and H.~Kashima, ``Estimation of low-rank tensors via
  convex optimization,'' \emph{\url{http://arxiv.org/abs/1010.0789}}, 2010.

\bibitem{33}
{Intel Berkeley dataset},
  Available:\url{http://http://db.lcs.mit.edu/labdata/labdata.html/}.

\bibitem{24}
A.~Smoli¨½ski, B.~Walczak, and J.~W. Einax, ``Exploratory analysis of data sets
  with missing elements and outliers,'' \emph{Chemosphere}, vol.~49, no.~3, pp.
  233--245, 2002.

\end{thebibliography}
\end{document}